\documentclass[twocolumn]{revtex4}

\usepackage{amsmath,amsfonts,amssymb}
\usepackage[english]{babel}
\usepackage[latin1]{inputenc}
\usepackage[T1]{fontenc}
\usepackage{color}
\usepackage{float}
\usepackage{verbatim}
\usepackage{graphicx}
\usepackage{bm}
\usepackage{mathtools}
\usepackage{stmaryrd}
\usepackage{anyfontsize}

\usepackage[font={small}]{caption}
\usepackage{subcaption}
\usepackage{color}

\newcommand{\beginsupplement}{%
        \clearpage
        \setcounter{table}{0}
        \renewcommand{\thetable}{S\arabic{table}}%
        \setcounter{figure}{0}
        \renewcommand{\thefigure}{S\arabic{figure}}%
     }

\begin{document}

\title{Eco-evolutionary dynamics and collective dispersal: implications for salmon metapopulation robustness}
\author{
Justin D. Yeakel${}^{1,2,*}$, Jean P. Gibert${}^{1}$, Peter A. H. Westley${}^{3}$, \& Jonathan W. Moore${}^{4}$ \\
${}^1$School of Natural Sciences, University of California, Merced, Merced CA, USA \\
${}^2$The Santa Fe Institute, Santa Fe NM, USA \\
${}^3$College of Fisheries and Ocean Sciences, University of Alaska, Fairbanks, Fairbanks AK, USA \\
${}^4$Earth${}_2$Oceans Research Group, Simon Fraser University, Burnaby BC, Canada \\
${}^*$To whom correspondence should be addressed: jdyeakel@gmail.com
}

\begin{abstract} 
The spatial dispersal of individuals is known to play an important role in the dynamics of populations, and is central to metapopulation theory. At the same time, local adaptation to environmental conditions creates a geographic mosaic of evolutionary forces, where the combined drivers of selection and gene flow interact. Although the dispersal of individuals from donor to recipient populations provides connections within the metapopulation, promoting demographic and evolutionary rescue, it may also introduce maladapted individuals into habitats host to different environmental conditions, potentially lowering the fitness of the recipient population. Thus, dispersal plays a dual role in both promoting and inhibiting local adaptation. Here we explore a model of the eco-evolutionary dynamics between two populations connected by dispersal, where the productivity of each is defined by a trait complex that is subject to local selection. Although general in nature, our model is inspired by salmon metapopulations, where dispersal between populations is defined in terms of the straying rate, which has been shown to be density-dependent, and recently proposed to be shaped by social interactions consistent with collective movement. The results of our model reveal that increased straying between evolving populations leads to alternative stable states, which has large and nonlinear effects on two measures of metapopulation robustness: the portfolio effect and the time to recovery following an induced disturbance. We show that intermediate levels of straying result in large gains in robustness, and that increased habitat heterogeneity promotes robustness when straying rates are low, and erodes robustness when straying rates are high. Finally, we show that density-dependent straying promotes robustness, particularly when the aggregate biomass is low and straying is correspondingly high, which has important ramifications for the conservation of salmon metapopulations facing both natural and anthropogenic disturbances.
\end{abstract}

\maketitle

\vspace{2mm}
\noindent {\bf Media Summary} 
Many migratory species, such as salmon, are remarkable in finding their way home. This homing has allowed fine-scale adaptations to the environments in which they evolve. But some individuals do not find their way home and instead stray to other locations, especially when there are fewer individuals to help with collective decision-making. With an eco-evolutionary model, we discovered that an intermediate and density-dependent straying rate allows linked populations to be robust to disturbance but maintain local adaptations.\\

\section{Introduction}

Intraspecific diversity can increase the resilience and stability of species or metapopulations. 
This diversity-stability linkage occurs when there are asynchronous population dynamics, where the changes in population size varies temporally across the metapopulation. 
This asynchrony will increase the potential for demographic rescue \citep{Brown:1977gk,Earn:2000fm} and also decrease the variability of processes that integrate across the metapopulation \citep{Anonymous:2015gf}. 
For example, different responses to climate variability within populations of a rare plant reduced fluctuations in abundance \citep{Abbott:2017hl}. 
This statistical buffer has traditionally been quantified as the Portfolio Effect (PE), which is the ratio of the population CV to the CV of the aggregated metapopulation \citep{Thibaut:2012km}. 
Strengthened portfolio effects are expected to increase the robustness of metapopulations to external disturbances, and by extension promote persistence \citep{Thibaut:2012km}.
In contrast, homogenization of populations leading to greater synchronization and weakened PE may be a harbinger of metapopulation collapse and extinction.

Movement of individuals among local populations (i.e. dispersal) can have a large influence on metapopulation persistence \citep{MilnerGulland:2011vm}. 
Dispersal facilitates evolutionary rescue, whereby immigration of individuals with heritable adapative traits can rescue small populations from local extinction in the context of maladaptive environmental change \citep{Bell:2011ki,Carlson:2014is}.
On the other hand, high rates of dispersal may synchronize populations and actually increase the risk of extinction of the entire metapopulation \citep{Earn:2000fm}. 
Dispersal will also influence the evolutionary dynamics of the metapopulation.
Although the dispersal of individuals into sites hosting other populations provides connections within the larger metapopulation, potentially promoting demographic and evolutionary rescue, it may also introduce maladapted individuals into habitats that are host to different environmental conditions, possibly lowering the mean fitness of the recipient population \citep{Muhlfeld:2014hs}. 
More broadly, dispersal can provide a mechanism by which phenotypes are sorted in space rather than time and facilitates the spread of maladaptive genes \citep{Lowe:2015ft}.
Dispersal in this case may lead to genetic homogenization that erodes the asynchrony underpinning portfolio effects and metapopulation persistence. 

\begin{figure}
  \captionsetup{justification=raggedright,
singlelinecheck=false
}
\centering
\includegraphics[width=0.45\textwidth]{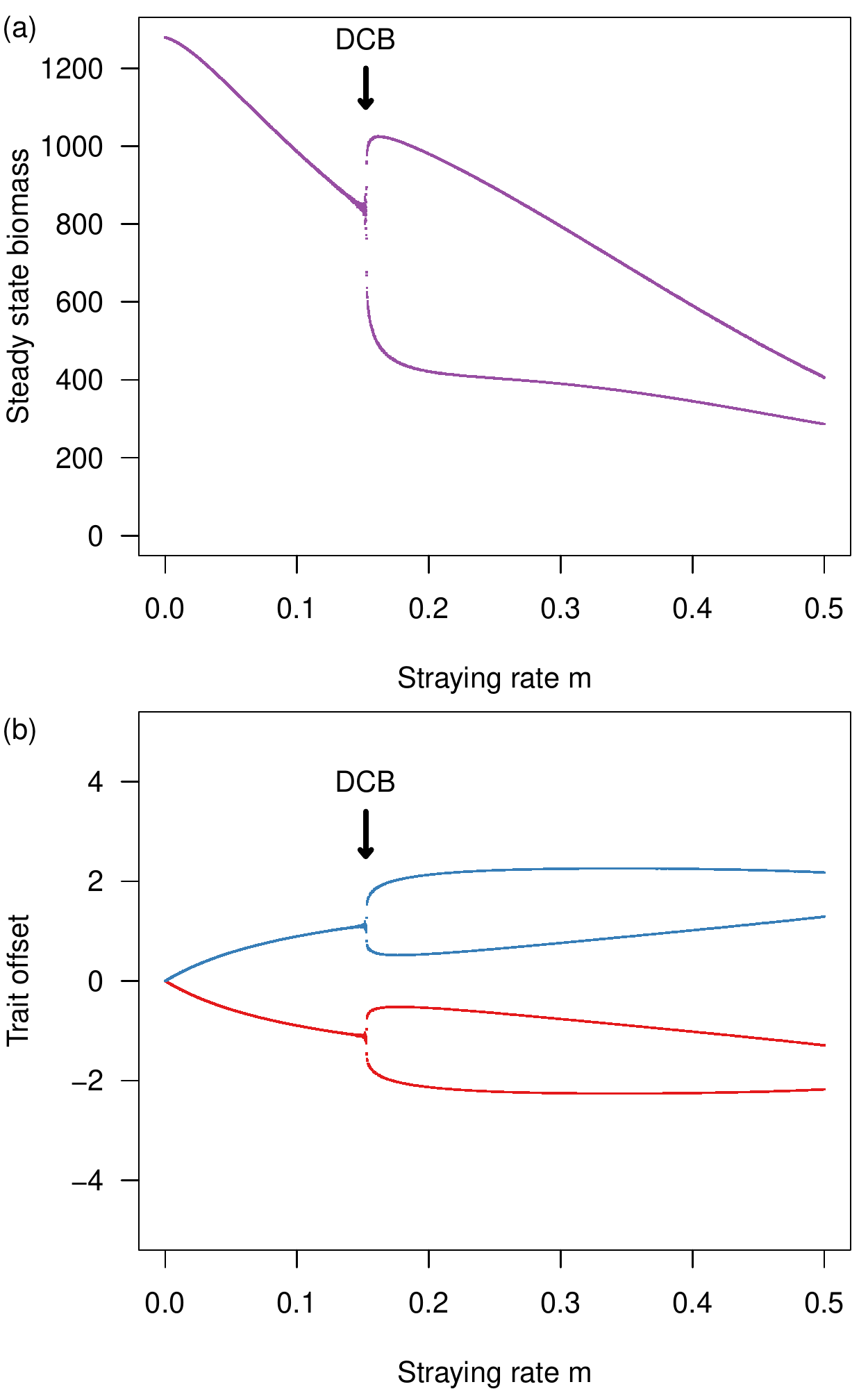}
\caption{
(a) The steady-state densities of $N_1$ and $N_2$ vs. the stray rate $m$. Which population attains the low- or high-density state is random due to small applied fluctuations in the initial conditions.
(b) The steady-state trait values measured as the offset from the local optimum $\theta_i - mu_i$, vs. the stray rate $m$. 
DCB marks the discrete cusp bifurcation.
Unless otherwise indicated, the default parameter values used are: $r_{\rm max}=2$; $Z=0.5$; $\beta=0.001$; $\theta_1=5$; $\Delta\theta=5$; $\tau=1$; $\sigma=1$; $T=1\times10^5$.
} \label{fig:traj}
\end{figure}

There is growing appreciation that a combination of abiotic, biotic, and anthropogenic factors can control the rate of dispersal among populations \citep{H:2013fs,Keefer:2014gg,Bett:2017ha}.
Migratory populations that return to breeding sites for reproduction are linked to each other by some proportion of the population that permanently disperses into the `wrong' site. 
Recently, the role of social interactions and collective navigation has been hypothesized \citep[][this volume]{Berdahl:2015kv,Berdahl:2016dx,HardestyMoore:wg}.
The rate at which individuals disperse may be linked to errors made at an individual-level that are themselves diminished by migrating in groups and pooling individual choices \citep{Simons:2004jo,Berdahl:2015kv,Berdahl:2016dx}.
The potential influence of collective dispersal on the dynamics of individual populations and the metapopulation as a whole is a topic of considerable interest that has tangible conservation implications \citep{Brenner:2012gl,Johnson:2012fe,Fullerton:2011ii}.


The eco-evolutionary impacts of dispersal likely have important implications for conservation and management in key taxa such as in migratory salmon.
While anadromous salmonid fishes (genera \emph{Oncorhynchus} and \emph{Salmo}) are renown for returning to their natal spawning habitats with high accuracy and precision after years at sea \citep{Quinn:2011tf,Jonsson:2011kg,Keefer:2014gg}, there are generally some individuals that `stray' (synonymously used hereafter to refer to dispersal) to non-natal sites to spawn \citep{Quinn:1993ge,Hendry:2004wf}.
Salmon may operate as metapopulations, where populations are genetically distinct but linked by some level of straying \citep{Schtickzelle:2007wb,Anderson:2014cx}.
Although extensive work has been done to document the extent of straying from donor populations into recipient populations \citep{Keefer:2014gg,Bett:2017ha}, only recently have the abiotic, biotic, and anthropogenic influences of straying behaviors been investigated systemically \citep{Keefer:2008bs,Westley:2015to,Bond:2016dz}.
Straying among salmon may be influenced by environmental factors such as water temperature, human activities such as hatchery practices, and population density as predicted by the collective navigation hypothesis \citep{Peterson:2014gy,Berdahl:2017uu}.
Straying can introduce new maladaptive genotypes into the recipient population, while the ensuing genetic homogenization could synchronize population dynamics and erode portfolio effects \citep{Moore:2010gs,Carlson:2011ce,Braun:2016ib}.
Thus, there is an opportunity and need to consider the eco-evolutionary consequences of straying for metapopulations in species of conservation and management concern such as salmon. 

Here we seek to explore how collective density-dependent straying influences the stability and robustness of metapopulations through ecological and evolutionary processes.
To address this question we constructed a eco-evolutionary model of two populations occupying different sites that are linked by straying individuals, each with an associated trait distribution subject to natural selection determined by local conditions.
Specifically we compared (a) different rates of straying and (b) the influence of collective movement, across (c) increasing environmental heterogeneity, by assessing two measures of metapopulation robustness: the portfolio effect and the time required for a population(s) to recover following an induced disturbance. 
This model enables us to explore the tradeoff between the potentially detrimental erosion of local adaptation vs. the positive effects of demographic and evolutionary rescue, both of which are facilitated by straying. 


\begin{figure*}
  \captionsetup{justification=raggedright,
singlelinecheck=false
}
\centering
\includegraphics[width=1\textwidth]{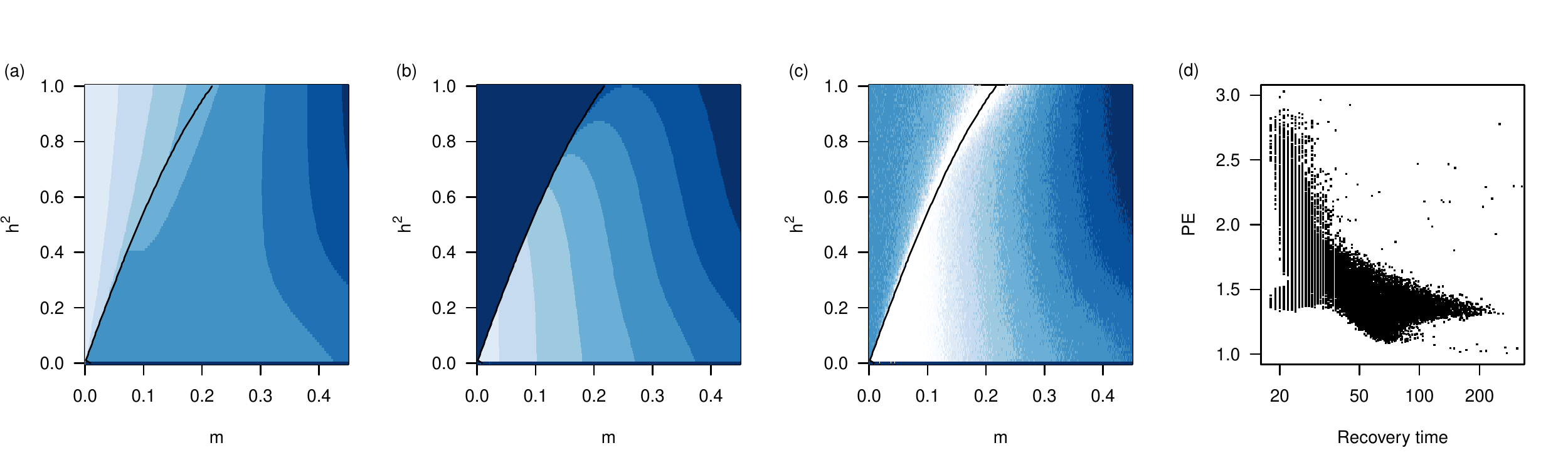}
\caption{
(a) Total means $N_t^*$, 
(b) difference in means $\Delta N^*$, and 
(c) the portfolio effect PE as a function of heritability $h^2$ and a constant stray rate $m$. Light colors = high values.
The black line shows the cusp bifurcation separating a single steady-state (left) from alternative stable states (right).
(d) The relationship between the time to recovery following a disturbance and the portfolio effect. Portfolio effects greater than unity corresponds to less synchronization.
} \label{fig:PE}
\end{figure*}

\section{Model Description \& Analysis}

\noindent{\bf (a) Metapopulation framework}\\
\noindent We considered two populations $N_1$ and $N_2$ that inhabit two distinct habitats, each with trait values $x_1$ and $x_2$ determining recruitment rates.
We assumed that there is an optimum trait value $\theta_1$ and $\theta_2$ associated with each habitat, where recruitment is maximized if the trait value of the local population equals the optimum, such that $x = \theta$.
Moreover, we assumed that $x_{1,2}$ are normally distributed with means $\mu_1$ and $\mu_2$ and have the same standard deviation $\sigma$.
As such, the recruitment rate $R_1[\mu_1(t),\theta_1]$ for both populations is determined by the mean trait value of the local population relative to optimal value at that site.
Trait means for each population are subject to selection, the strength of which is proportional to the difference between the trait mean and the local trait optimum at a given point in time \citep{simpson1953major,Lande:1976ga}.

The two populations occur in spatially separate sites that are close enough that a proportion of the population $m$ can stray into the other site, and where mortality occurs before individuals return to reproduce.
If there is no straying between these populations, then the mean trait evolves towards the optimal value for each site $\mu_1 \rightarrow \theta_1$, and the recruitment rate for each population is maximized.
If there is straying between populations at rate $m$, then the traits in each respective location will be pulled away from the optimum, and recruitment rates will be lowered.
As $m \rightarrow 0.5$, the populations are perfectly mixed, acting as a single population.

We used the discrete Ricker framework described by Shelton and Mangel \citep{Shelton:2011eq} as the basis for our two-site metapopulation model, with the added effect of the local population $N_i$ mixing with a set proportion $m$ of a remote population $N_j$ that is straying into it.
In this sense, both populations serve simultaneously as donor and recipient populations.
We first assumed that a proportion ${\rm e}^{-Z}$ of both populations survive such that the surviving aggregated population, composed of both local individuals (at site $i$) and incoming strays (from site $j$), is $\left((1-m)N_i(t) + m N_j(t) \right){\rm e}^{-Z}$.
Because local individuals will recruit differently than incoming strays, the recruitment of the aggregate must incorporate two recruitment functions, given by $\left(R_i[\mu_i(t)] (1-m)N_i(t) + R_i[\mu_j(t)] m N_j(t)\right)$.
This mix of individuals is subject to identical compensatory effects, which is determined by the parameter $\beta$.
Taken together, the difference equation that determines changes in population size is

\begin{align}
  &N_i(t+1) = \\ \nonumber
  &\left((1-m)N_i(t) + m N_j(t) \right){\rm e}^{-Z} \\ \nonumber
  &+ \left(R_i[\mu_i(t)] (1-m)N_i(t) + R_i[\mu_j(t)] m N_j(t)\right) \\ \nonumber
  &\times {\rm e}^{-\beta ((1-m)N_i(t) + m N_j(t))},
  \label{eq:N}
\end{align}

\noindent where the equation for $N_j$ mirrors that for $N_i$.

The recruitment of local individuals $(1-m)N_i(t)$ as a function of their mean trait value at time $t$ and the local trait optimum, is

\begin{align}
  &R_i[\mu_i(t)] = \\ \nonumber
  &\int_{-\infty}^\infty r_{\rm max}\exp\left\{\frac{(x_i-\theta_i)^2}{2\tau^2}\right\} {\rm pr}(x_i,\mu_i(t),\sigma^2) {\rm d}x_i +\tilde{P}_i\\ \nonumber
  &= \frac{r_{\rm max} \tau  }{\sqrt{\sigma ^2+\tau ^2}}\exp\left\{-\frac{(\theta_i-\mu_i(t))^2}{2 \left(\sigma ^2+\tau ^2\right)}\right\} +\tilde{P}_i,
  \label{eq:R}
\end{align}

\noindent where the mismatch between the local trait mean $\mu_i(t)$ and the local optimum $\theta_i$ scales the recruitment rate for the population, and $\tilde{P}_i\sim {\rm Normal}(0,0.01)$ introduces a small amount of demographic error.
The parameter $\tau$ is the strength of selection, and controls the sensitivity of recruitment to changes in the mean trait value away from the optimum (the strength of selection increases with smaller values of $\tau$), which we set as $\tau=1$ here and throughout.
Because straying individuals are emigrating from a population with a mean trait value farther from the local optimum, their rate of recruitment is diminished.
Recent studies of wild sockeye salmon have indeed found that straying individuals have lower life-time fitness than individuals that do not stray, although it is unknown at what life-stage this selection occurs \citep{Peterson:2014gy}.
\\

Because individuals from the local population are mixed with individuals from the remote population via straying and subsequent reproduction, the resulting trait distribution is a mixed normal with weights corresponding to the proportion of the mixed population that are local individuals, $w_i$, and straying individuals, $1-w_i$, where 
\begin{equation}
w_i=\frac{(1-m)N_i(t)}{(1-m) N_i(t) + m N_j(t)}.
\end{equation}
We made two simplifying assumptions.
First, we assumed that the distribution resulting from the mix of remote and local individuals, following reproduction, is also normal with a mean value equal to that of the mean for the mixed-normal distribution.
Thus, strays can successfully reproduce and introduce their genotypes into the recipient population, which is supported by observations in wild populations \citep{Jasper:2013cc}.
Second, we assumed that changes in trait variance through time are minimal, such that $\sigma^2$ is constant over time, which is a common simplification in eco-evolutionary models of population dynamics \citep{Lande:1976ga,Schreiber:2011wx,Gilbert:2014ee,Gibert:2015kc}.

Following Lande \citep{Lande:1976ga}, the mean trait value thus changes through time according to the difference equation

\begin{align}
  \label{eq:mu}
  \mu_i(t+1) &= w_i\mu_i(t) + (1-w_i)\mu_j(t) \\ \nonumber
  &+ h^2\sigma^2\frac{\partial}{\partial \mu_i}\ln\left(w_i R_i[\mu_i(t)] + (1-w_i)R_i[\mu_j(t)]  \right),
\end{align}

\noindent where the first two components determine the mixed normal average of the aggregated local and remote populations.
The partial derivative in Eq. \ref{eq:mu} determines how the mean trait changes through time due to natural selection \citep{Lande:1976ga}, which is proportional to the change in mean fitness with respect to $\mu_i$.
We note that the derivative is dependent on both $\mu_i$ and $\mu_j$ due to the influence of the logarithm on the sum.
This model formulation has parallels to that proposed by Ronce and Kirkpatrick \citep{Ronce:2001dp}, where habitat specialization evolves between two populations as a function of dispersal, yet differs in that we treat trait evolution mechanistically at some cost to analytical tractability.
\\

\begin{table}
\begin{center}
  \caption{Parameters and definitions}
\begin{tabular}{ l l }
\hline
Parameter & Definition \\
\hline
$N_i(t)$, $N_T(t)$ & Individual, aggregate population over time\\
$x_i$ & Trait value for an individual in population $i$\\
$\mu_i(t)$ & Mean of $x$ for population $i$ over time\\
$\sigma^2$ & Genetic variance of trait $x$\\
$m$, $m(t)$ & Constant, density-dependent straying rate\\
$m_0$ & Straying rate of an individual\\
$R_i[\mu(t)]$ & Recruitment rate of population $i$\\
$r_{\rm max}$ & Maximum recruitment rate\\
${\rm e}^{-Z}$ & Survival rate\\
$\beta$ & Strength of density dependence\\
$\theta_i$ & Optimal trait value for habitat $i$\\
$\Delta\theta$ & Habitat heterogeneity\\
$\tau$ & Strength of selection\\
$h^2$ & Heritability\\
$C$ & Half saturation constant for $m(t)$\\
${\rm PE}$ & Portfolio Effect\\
\hline
\end{tabular}
\end{center}
\end{table}

\noindent {\bf (b) Density-dependent straying}
\noindent We have so far assumed that the proportion of strays leaving and entering a population is constant, however there is mounting evidence that at least in some species (including salmon) the straying rate is density-dependent, a signature of collective navigation \citep{Berdahl:2014bl,Berdahl:2017uu}.
Specifically, the rate at which individuals stray has been linked directly to a collective decision-making phenomenon, where greater numbers of individuals tend to decrease the rate at which individuals err, reducing the overall proportion of a population that strays.
According to Berdahl et al. \citep{Berdahl:2016dx}, given the probability that an individual strays is $m_0$, the proportion of the local population $N_i(t)$ that strays is

\begin{equation}
  m(t) = m_0\left(1- \frac{N_i(t)}{C+N_i(t)}\right),
  \label{eq:ddm}
\end{equation}

\noindent where $C$ is a half-saturation constant and is set to $C=1000$ throughout.
When the population density is very high, $m(t) \rightarrow 0$, and when the population is small, individuals operate without regard to collective behavior, such that $m(t) \rightarrow m_0$.
For realistic population densities, $m(t) < m_0$.\\

\begin{figure*}
  \captionsetup{justification=raggedright,
singlelinecheck=false
}
\centering
\includegraphics[width=0.9\textwidth]{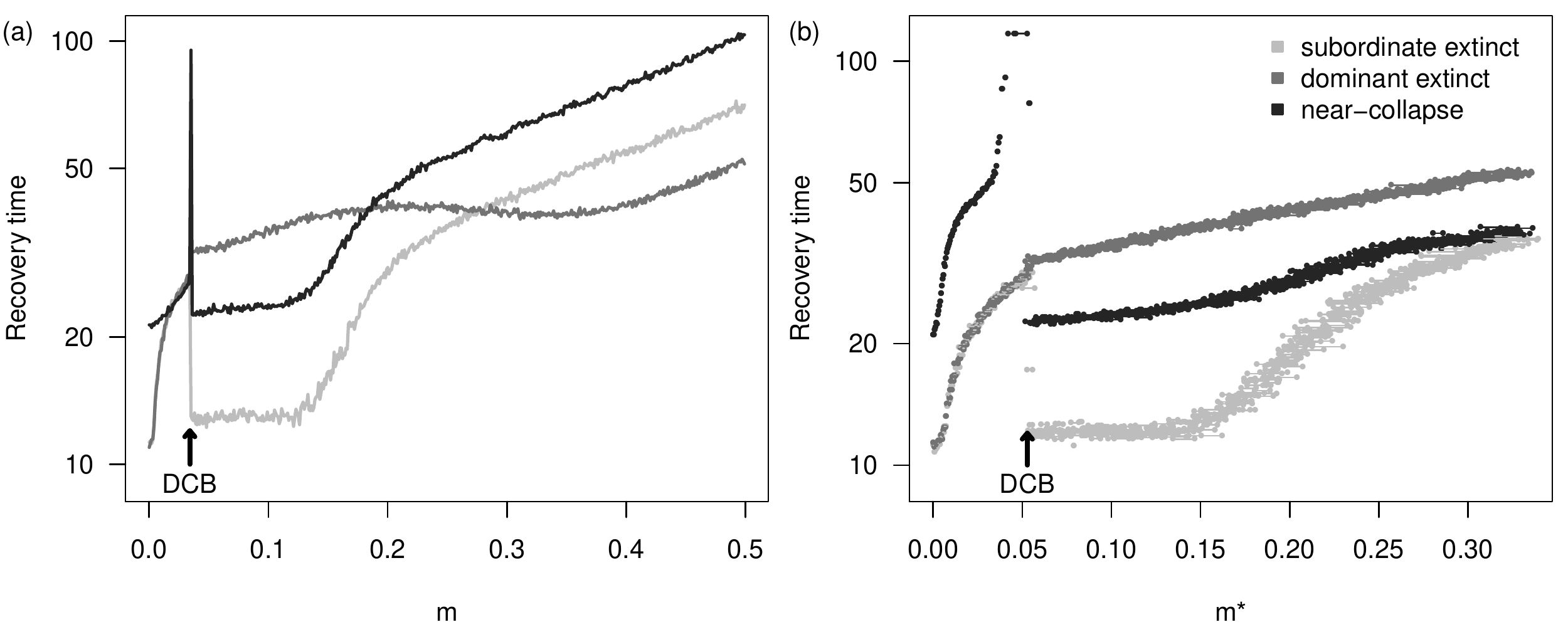}
\caption{
Recovery time of $N_T^*$ following the extinction of either the low-density (light gray) or high-density (gray) population, or the near-collapse of both (dark gray) assuming (a) constant straying rates $m$ and (b) density-dependent straying rates (evaluated at the steady-state $m^*$) with trait heritability $h^2=0.2$.
If $m$ is density-dependent, in the alternative stable state regime there are two straying rates observed: one each for the low- and high-density populations, respectively, which are linked by a horizontal line.
DCB marks the discrete cusp bifurcation.
} \label{fig:relax}
\end{figure*}

\noindent {\bf (c) Habitat heterogeneity}
\noindent Increasing differences in optimal trait values between sites ($\Delta\theta = \left|\theta_i - \theta_j\right|$) corresponds to greater regional differences in the conditions that favor alternative trait complexes, which can be interpreted as increased habitat heterogeneity.
If both populations are isolated, natural selection will direct the mean trait values of both populations towards their respective optima, such that $\mu_i(t) \rightarrow \theta_i$ as $t\rightarrow\infty$.
Habitat heterogeneity and the rate of straying are treated both independently, and as parameters that covary.
In the latter instance, we examined a case where it is assumed that increased habitat heterogeneity correlates with lower straying rates, and vice versa (illustrated in figure \ref{fig:mthetarelation}).
Two scenarios may lead to this correlation: 
(\emph{i}) sites may be distributed over greater spatial distances, where habitat differences are assumed to be exaggerated and the likelihood of straying over greater distances is lower \citep{Candy:2000hu,JPE:JPE1383};
(\emph{ii}) individuals may have behaviors promoting dispersal between habitats with structural or physiognomic similarities \citep{Peterson:2014gy}.
In this case, the rate of straying would be greater between habitats with smaller differences in trait optima (lower $\Delta\theta$) and lesser between habitats with greater differences in trait optima (higher $\Delta\theta$).
\\

\noindent {\bf (d) Measuring metapopulation robustness}
\noindent We evaluated metapopulation robustness by measuring the average-CV portfolio effect (PE) \citep{Anderson:2014cx,Schindler:2015gf} as well as the recovery time, which is the time required for the system to return to a steady-state following an induced disturbance to one or both of the populations \citep{Ovaskainen:2002il}.
Throughout, we refer to an increase in portfolio effects and/or reduction in recovery time as promoting metapopulation robustness.

The average-CV portfolio effect is, as the name implies, the average CV across each population $N_i$ divided by the CV of the aggregate $N_T=\sum_i N_i$ \citep{Anderson:2013gb}, such that

\begin{equation}
\langle{\rm PE}\rangle =\frac{1}{X}\sum_{i=1}^{X} \frac{\sqrt{{\rm VAR}(N_i^*)}}{{\rm E}(N_i^*)}\cdot \frac{{\rm E}(N_T^*)}{\sqrt{{\rm VAR}(N_T^*)}},
\label{eq:pe}
\end{equation}

\noindent where in this case the number of populations is limited to $X=2$ and the expectations $\rm E(\cdot)$ and variances $\rm VAR(\cdot)$ are evaluated at the steady-state.
The steady-state condition is denoted by `$*$'.
As the CV of $N_T^*$ decreases relative to that of the constituent populations, $\langle{\rm PE}\rangle > 1$, and the metapopulation is presumed to be more stable because the aggregate has functioned to dampen population-level variance.
Moreover, portfolio effects greater than unity correspond to less synchronization  \citep{Loreau:2008ju,Anderson:2014cx,Yeakel:2013vz} and thus a greater potential for demographic rescue among populations, buffering the system as a whole against extinction. 

\begin{figure*}
  \captionsetup{justification=raggedright,
singlelinecheck=false
}
\centering
\includegraphics[width=0.35\textwidth]{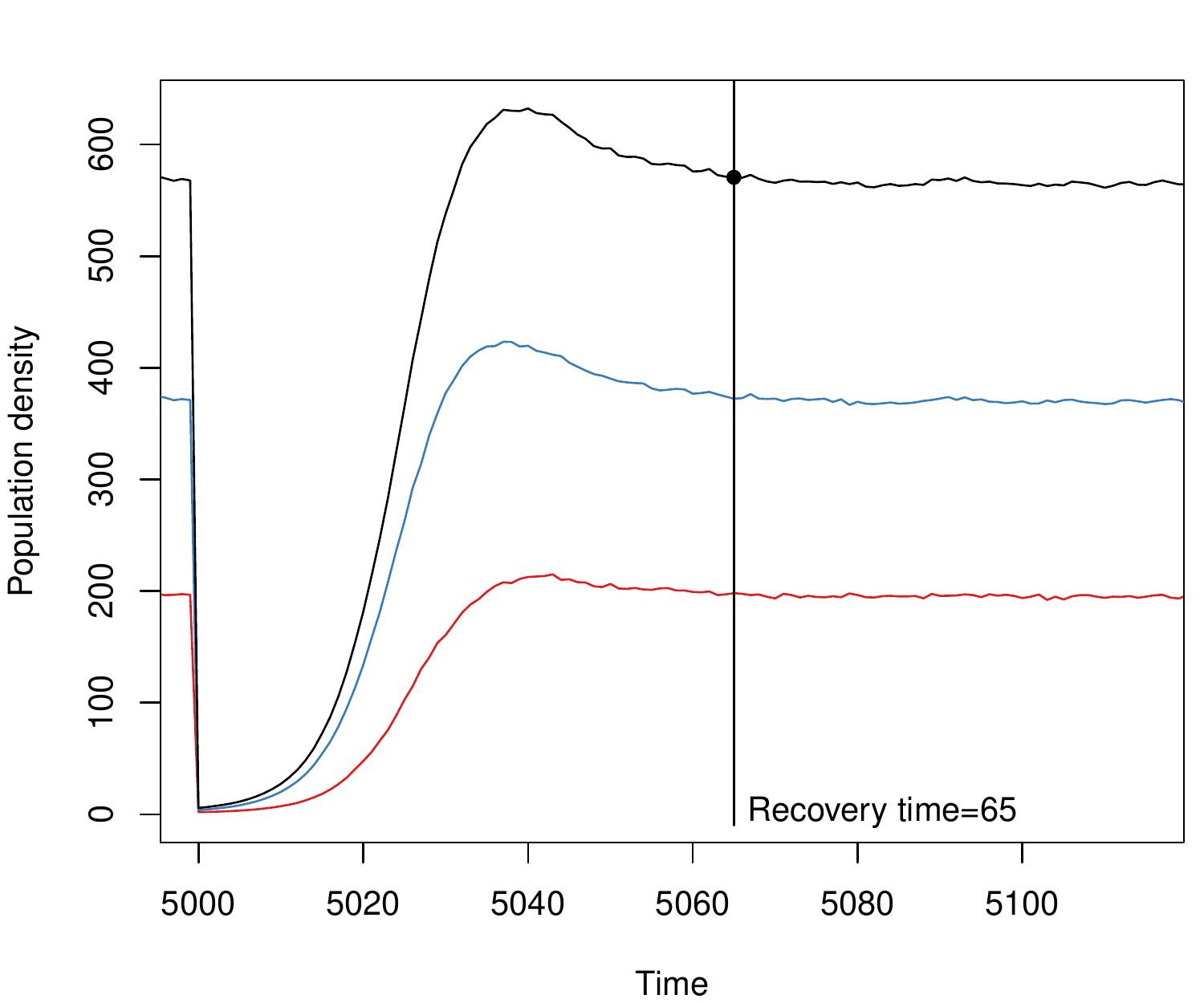}
\caption{
Example of the numerical procedure used to estimate recovery time. After a disturbance is introduced, the recovery time is calculated by measuring the point in time where $N_T$ (in black), which is the aggregate of both populations (blue, red) settles to within one standard deviation of the new equilibrium $N_T^*$. 
} \label{fig:recovery}
\end{figure*}

A more direct way to measure system robustness is to measure the time that the system (measured as the aggregate steady-state biomass $N_T^*$) takes to return to a steady-state following an induced disturbance: systems that recover quickly (shorter recovery times) are more robust than those that recover more slowly (longer recovery times).
Although there is a direct eigenvalue relationship between the rate of return following a small pulse perturbation \citep{GuckHolmes}, because we aimed to 
1) assess the effects of a large perturbation far from the steady-state, and 
2) estimate the time required for all transient effects to decay following this perturbation (including dampened oscillations), we used a simulation-based numerical procedure.
Recovery time was calculated by initiating a disturbance at $t=t_d$, and monitoring $N_T(t_d+t)$ as $t\rightarrow T$, where $T$ is large. 
The aggregate was deemed recovered at $t_r$, such that recovery time was calculated as $t_r-t_d$, and recovery at $t=t_r$ was measured as the initial $t$ where $N_T(t) < {\rm SD}\left( N_T^* \right)$ for $t\in(t_r,T)$, where $\rm{SD}(\cdot)$ is standard deviation (illustrated in figure \ref{fig:recovery}).

Numerically estimating the time that it takes for a perturbed system to recover also permits a more detailed perspective of metapopulation robustness.
For example, if populations settle to alternative stable states (one at low- and one at high-density), comparing recovery times after a disturbance applied to individual populations allows for an assessment of which component of the metapopulation has a longer-lasting influence on the system's recovery. 
We measured recovery time following three types of induced disturbance: (\emph{i}) extinction of the low-density population; (\emph{ii}) extinction of the high-density population (scenarios \emph{i} and \emph{ii} are equivalent if populations have the same density); (\emph{iii}) near-collapse of both populations where just 1.0\% of each survives.
\\

\begin{figure}
  \captionsetup{justification=raggedright,
singlelinecheck=false
}
\centering
\includegraphics[width=0.5\textwidth]{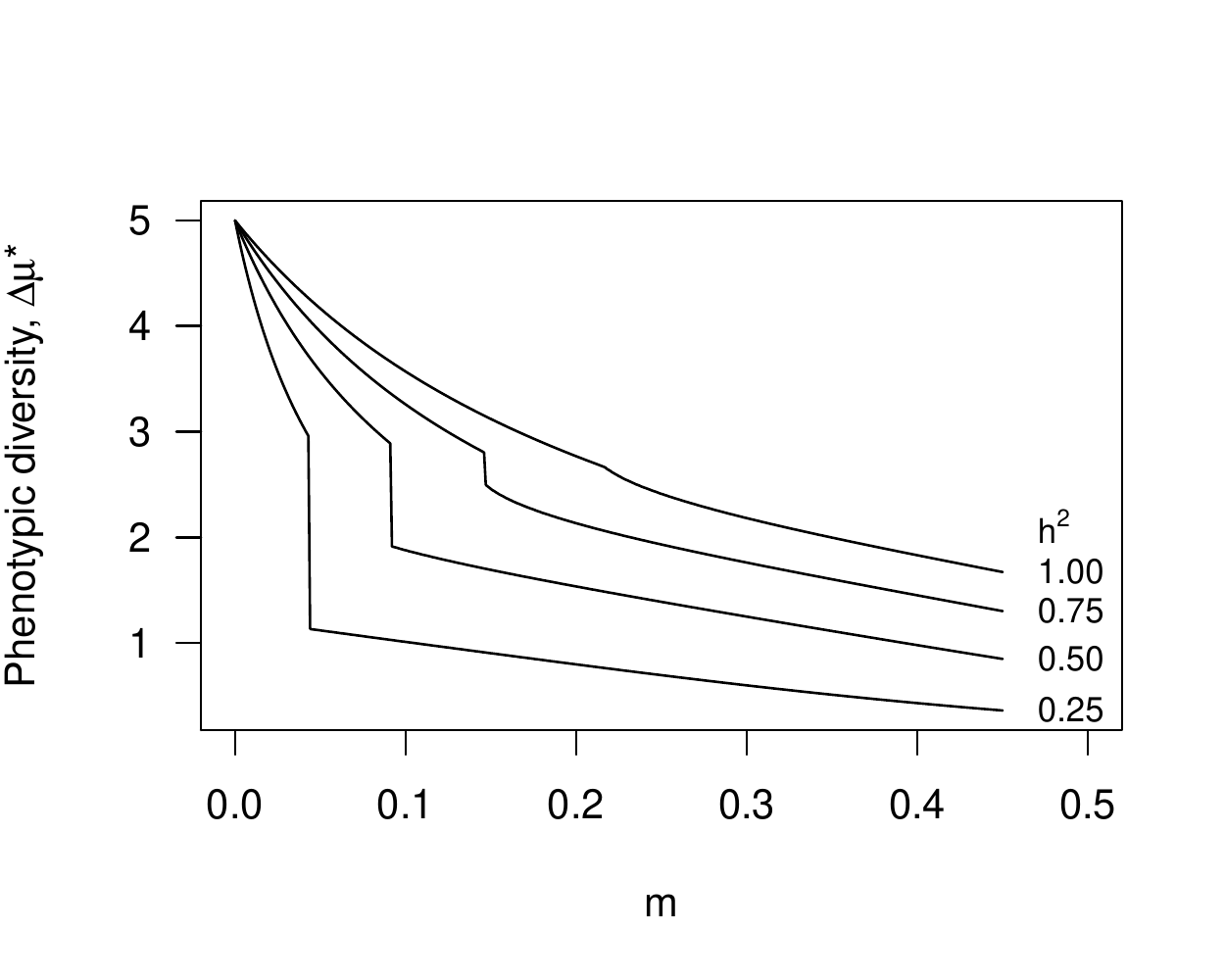}
\caption{
Phenotypic diversity ($\Delta \mu^*$) evaluated at the steady-state as a function of straying rate $m$ and trait heritability $h^2$. The jump occurs as the system crosses the cusp bifurcation; lower phenoytic diversity emerges with higher straying rates and in the alternative stable state regime. 
} \label{fig:traitdiff}
\end{figure}

\section{Results}

\begin{figure*}
  \captionsetup{justification=raggedright,
singlelinecheck=false
}
\centering
\includegraphics[width=0.35\textwidth]{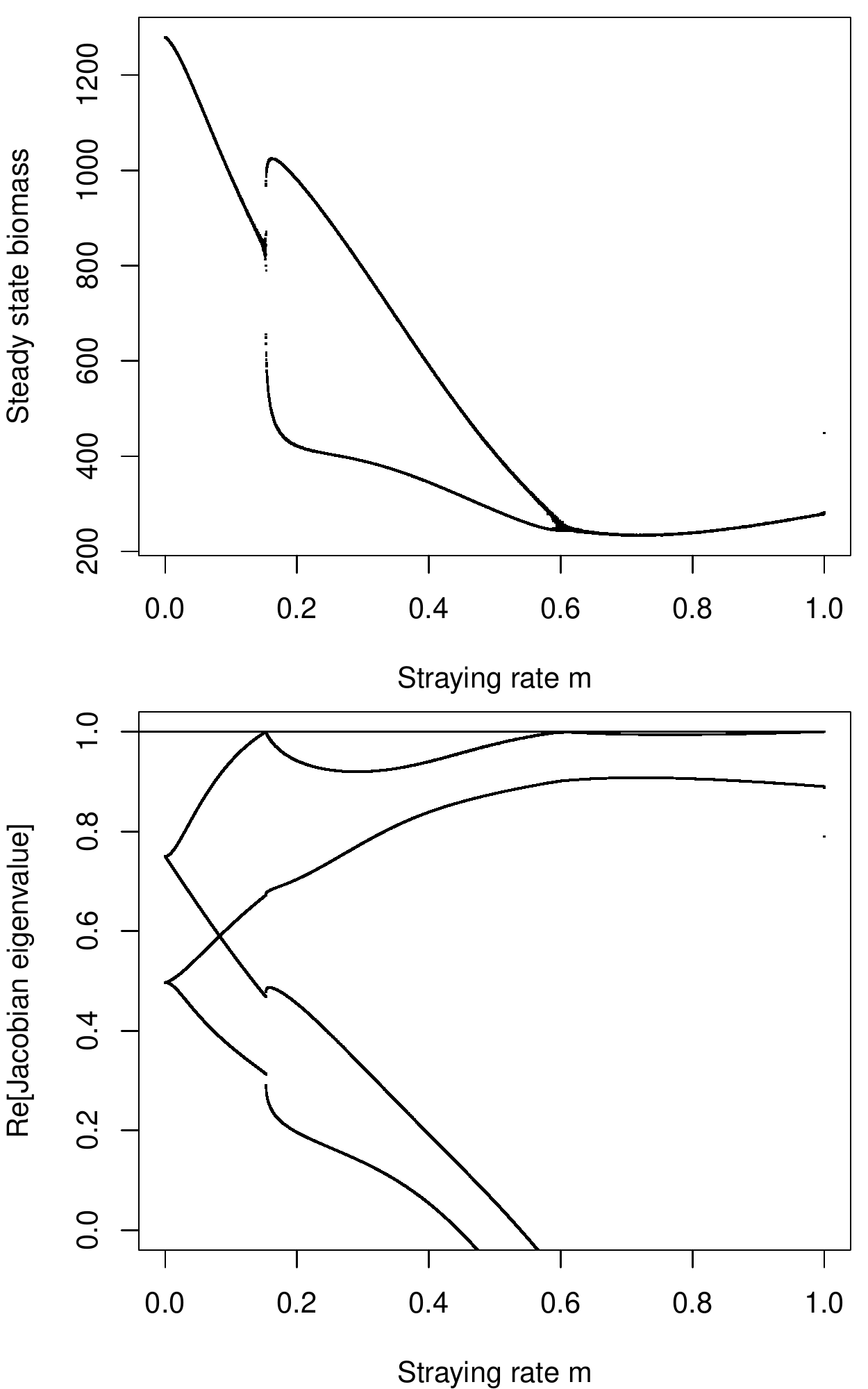}
\caption{
The real parts of the four eigenvalues for the Jacobian matrix of the 4-dimensional system.
The cusp bifurcation occurs when the dominant eigenvalue crosses the unit circle at $+1$. 
} \label{fig:eigs}
\end{figure*}

\noindent{\bf (a) Nonlinear effects of straying on metapopulation robustness} \\
\noindent Regardless of density dependence, straying lowers steady-state densities for both populations by (\emph{i}) the donor population losing locally-adapted individuals to the recipient population and (\emph{ii}) the introduction of maladapted individuals to the recipient population from the donor population (Fig. \ref{fig:traj}).
This prediction is in accordance with observations from natural populations \citep{Bett:2017ha}. 
The decline in steady-state densities is not gradual: as straying increases, the system crosses a discrete cusp bifurcation (DCB) \citep{AleksandrovichKuznetsov:1995p2580} whereby the single steady-state for the metapopulation bifurcates into two basins of attraction: one at high biomass, and one at low biomass density (figure \ref{fig:traj}a, \ref{fig:PE}a).
Mean trait values for both populations bifurcate similarly (figure \ref{fig:traj}b). 
In discrete systems, the cusp bifurcation is defined by two fold bifurcations intersecting at a cusp point \citep{AleksandrovichKuznetsov:1995p2580}, and is observed when the real part of the dominant eigenvalue of the Jacobian matrix crosses the unit circle at +1 (figure \ref{fig:eigs}).
Visually, the dynamics are similar to those observed at a pitchfork bifurcation in continuous systems, where a single steady-state gives rise to two alternative steady-states.

\begin{figure*}
  \captionsetup{justification=raggedright,
singlelinecheck=false
}
\centering
\includegraphics[width=0.35\textwidth]{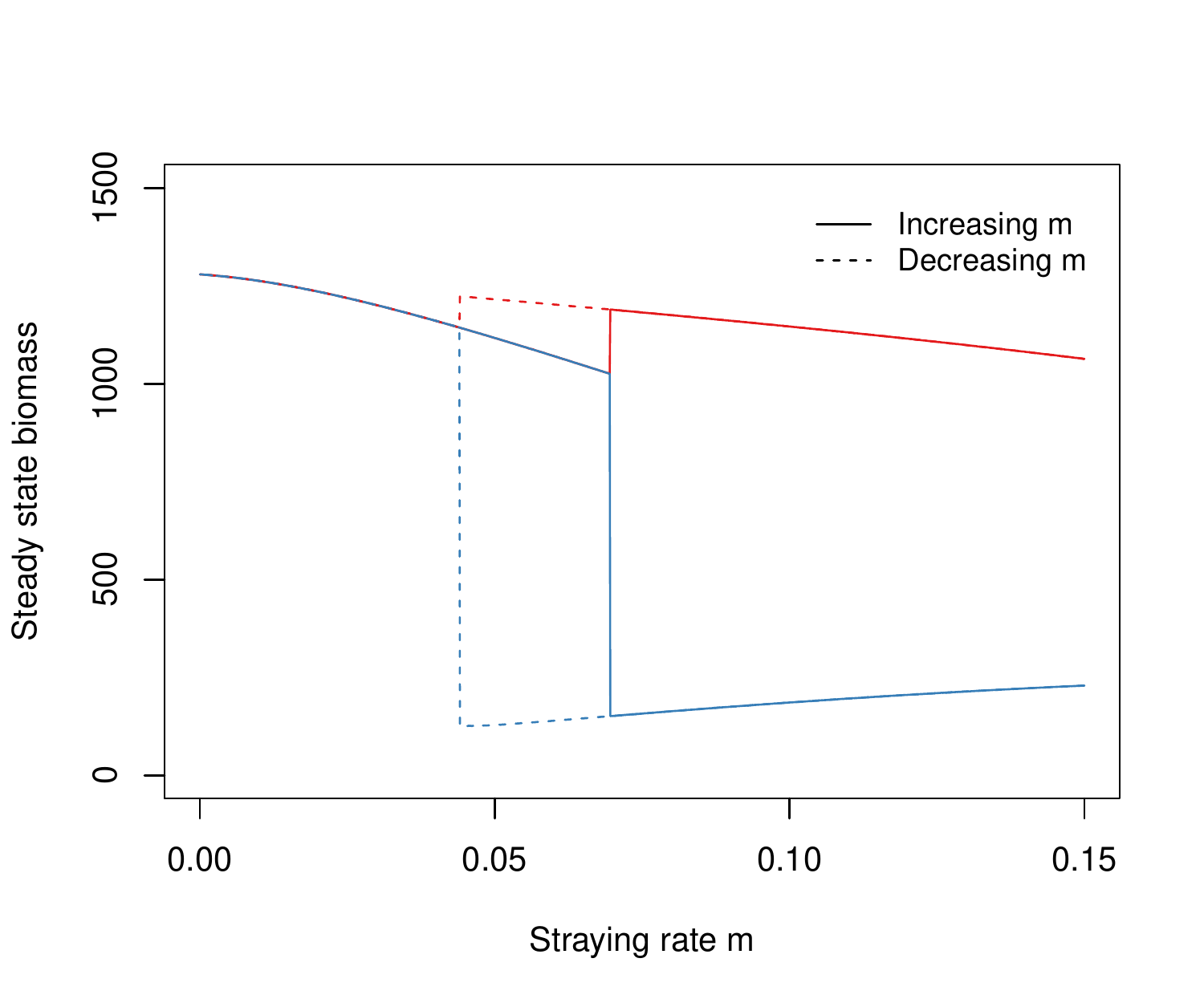}
\caption{
Increasing the straying rate results in the transition from a single steady-state for both populations to a dominant and subordinate states. If the straying rate is subsequently lowered, the single steady-state is not easily obtained, which is the hallmark of hysteresis.
} \label{fig:hysteresis}
\end{figure*}

Above the threshold straying rate defined by the DCB, there are two alternative eco-evolutionary states: the \emph{dominant state} population will have a higher density and greater degree of local adaptation (smaller trait offset from the local optimum), while the \emph{subordinate state} population will have lower density with maladapted trait values (larger trait offset from the local optimum). 
Hysteresis is observed to occur at this transition, such that the single steady-state regime cannot easily be recovered by reducing straying after the system attains alternative steady-states (figure \ref{fig:hysteresis}).
These dynamics are also observed in the Ronce and Kirkpatrick model, where populations are described as transitioning between symmetric to asymmetric states \citep{Ronce:2001dp}.
Whether a specific population goes to one state or the other in our model is random, due to a small amount of introduced variance in the initial conditions.

\begin{figure}
  \captionsetup{justification=raggedright,
singlelinecheck=false
}
\centering
\includegraphics[width=0.4\textwidth]{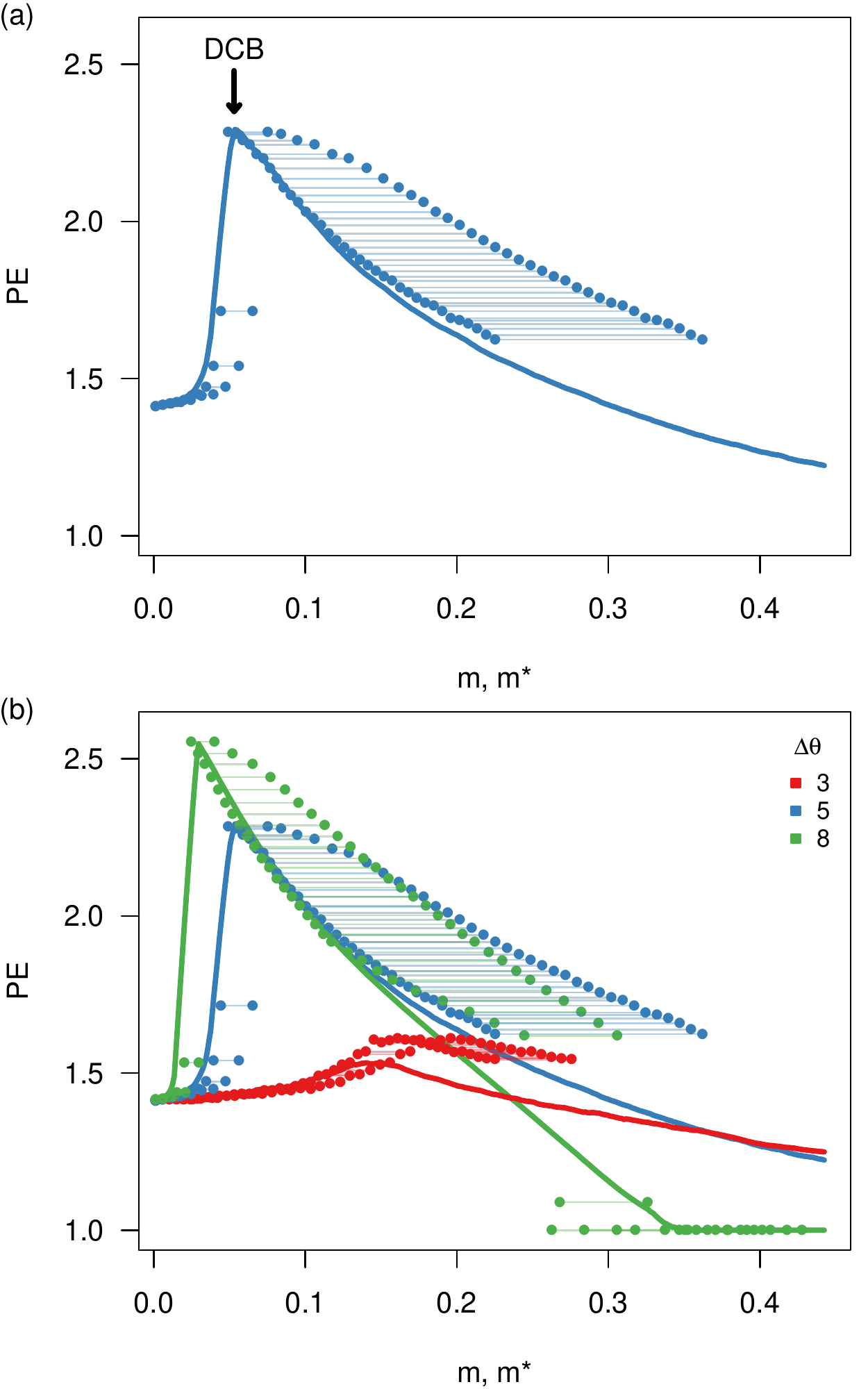}
\caption{
(a) Median portfolio effect as a function of a constant stray rate $m$ (solid line) and density-dependent stray rate (point pairs) given heritability is $h^2 < 0.5$ and $\Delta\theta=5$.
Point pairs connected by a horizontal line represent the PE as a function of density-dependent straying rates, evaluated for both low- and high-density populations at equilibrium. The lower straying rate of a pair is for the larger population; the higher straying rate is for the smaller population.
(b) Median portfolio effects for habitats with increasing heterogeneity as measured by the difference in regional trait optima $\Delta \theta$ for both constant and density-dependent stray rates as shown in (a).
Portfolio effects greater than unity corresponds to less synchronization.
DCB marks the discrete cusp bifurcation.
} \label{fig:thetaPE}
\end{figure}

Trait heritability $h^2$ has a large effect on the degree to which straying affects both the aggregate population steady-state density ($N^*_T$; figure \ref{fig:PE}a) as well as the difference between steady-state densities (the distance between alternative stable states: $\Delta N^*=|N^*_1-N^*_2|$; figure \ref{fig:PE}b).
Greater trait heritability results in a steeper decline in $N_T^*$ with increasing straying rates $m$, but leads to only moderate changes to $\Delta N^*$.
Lower trait heritability has little effect on the total biomass density but contrastingly large effects on $\Delta N^*$.
The cusp bifurcation (the black line in Figs. \ref{fig:PE}a-c) occurs at lower values of the straying rate $m$ with lower heritability (Fig \ref{fig:PE}a,b), indicating that weaker coupling between ecological and evolutionary dynamics in addition to higher rates of straying promotes the appearance of alternative stable states.
Although trait heritability among salmonids is variable, most life-history traits have an $h^2 <0.5$ \citep{Carlson:2008hl}, and we largely focus our efforts on that range.

As the cusp bifurcation is approached with increasing $m$, the portfolio effect increases sharply due to an amplification in variance within both donor and recipient populations.
This amplification in variance is the product of a dynamical process known as \emph{critical slowing down} that can occur near bifurcations \citep{Scheffer:2009gg}, a phenomenon that some have suggested may serve as an early warning indicator for approaching phase transitions \citep{Scheffer:2009gg,Lade:2012eu,Anonymous:2013br,Dakos:2014br,Krkosek:2014ch}.
For larger values of $m$ (to the right of the cusp bifurcation in Fig \ref{fig:PE}a-c), where alternative stable states occur, the portfolio effect declines steadily as the CV of $N_T^*$ increases.
The decline over $m$ is more gradual if trait heritability is low, and steeper if trait heritability is high (figure \ref{fig:PE}c).

As the portfolio effect is highly sensitive to the rate of straying between populations, so is the time required for the system to recover to a steady-state following a large disturbance.
In general, we find that the average-CV portfolio effect is negatively correlated with recovery time (figure \ref{fig:PE}d), indicating that, for our system, both measures are valuable indicators of metapopulation robustness.
Because we can assess the time to recovery in response to the various disturbance types described above, this allows us to gain an in-depth perspective into the fragility of the metapopulation as a function of straying rate.

\begin{figure*}
  \captionsetup{justification=raggedright,
singlelinecheck=false
}
\centering
\includegraphics[width=0.35\textwidth]{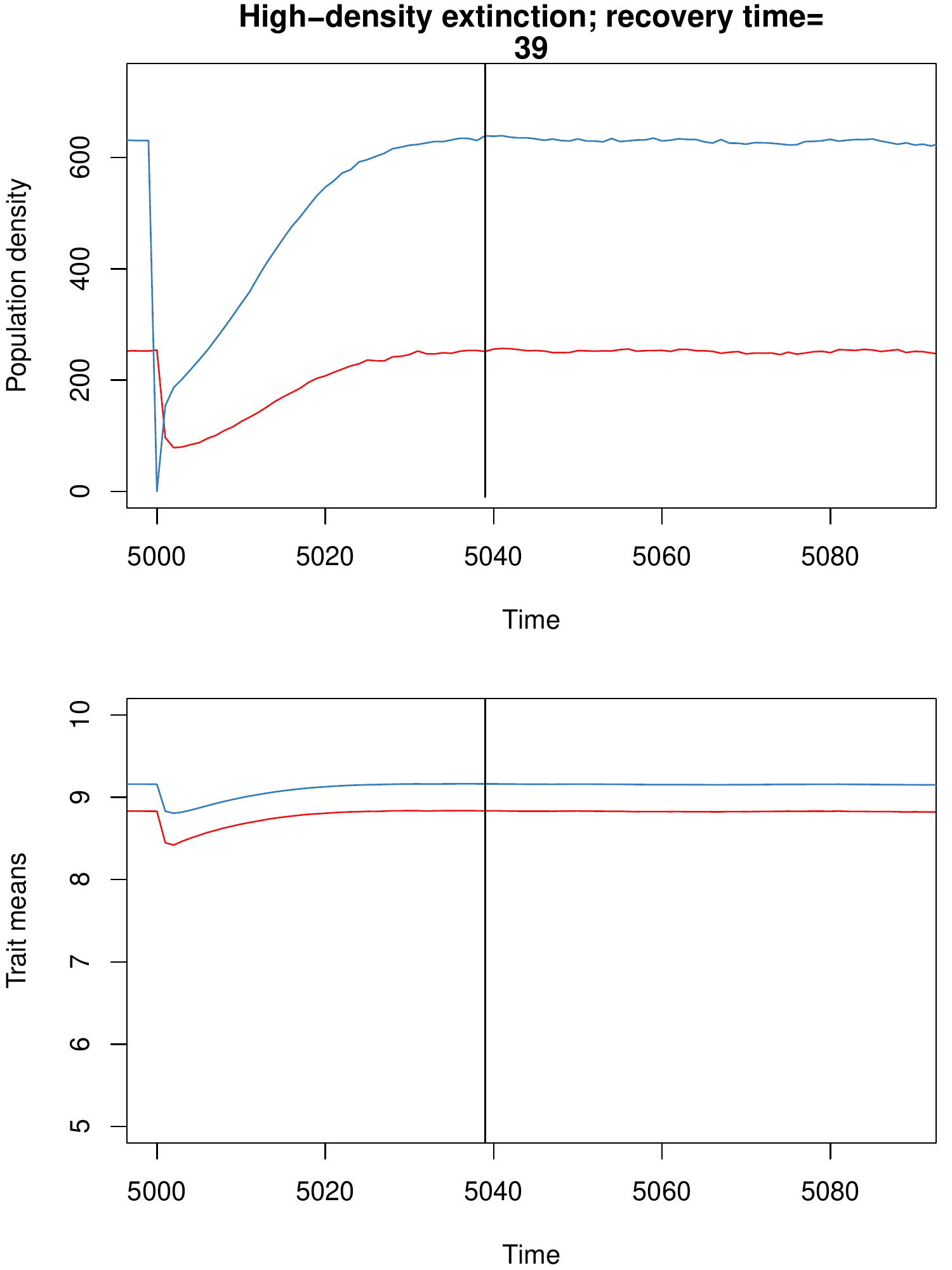}
\caption{
Extinction of high-density population with a high straying rate $m=0.4$ and low trait heritability $h^2=0.2$ (see figure \ref{fig:relax}a).
Black line marks the calculated point of recovery post-perturbation.
Trait optima are $\theta_1 = 10$ (blue population trajectory) and $\theta_2 = 5$ (red population).
} \label{fig:relaxtraj_hdlh}
\end{figure*}

\begin{figure*}
  \captionsetup{justification=raggedright,
singlelinecheck=false
}
\centering
\includegraphics[width=0.35\textwidth]{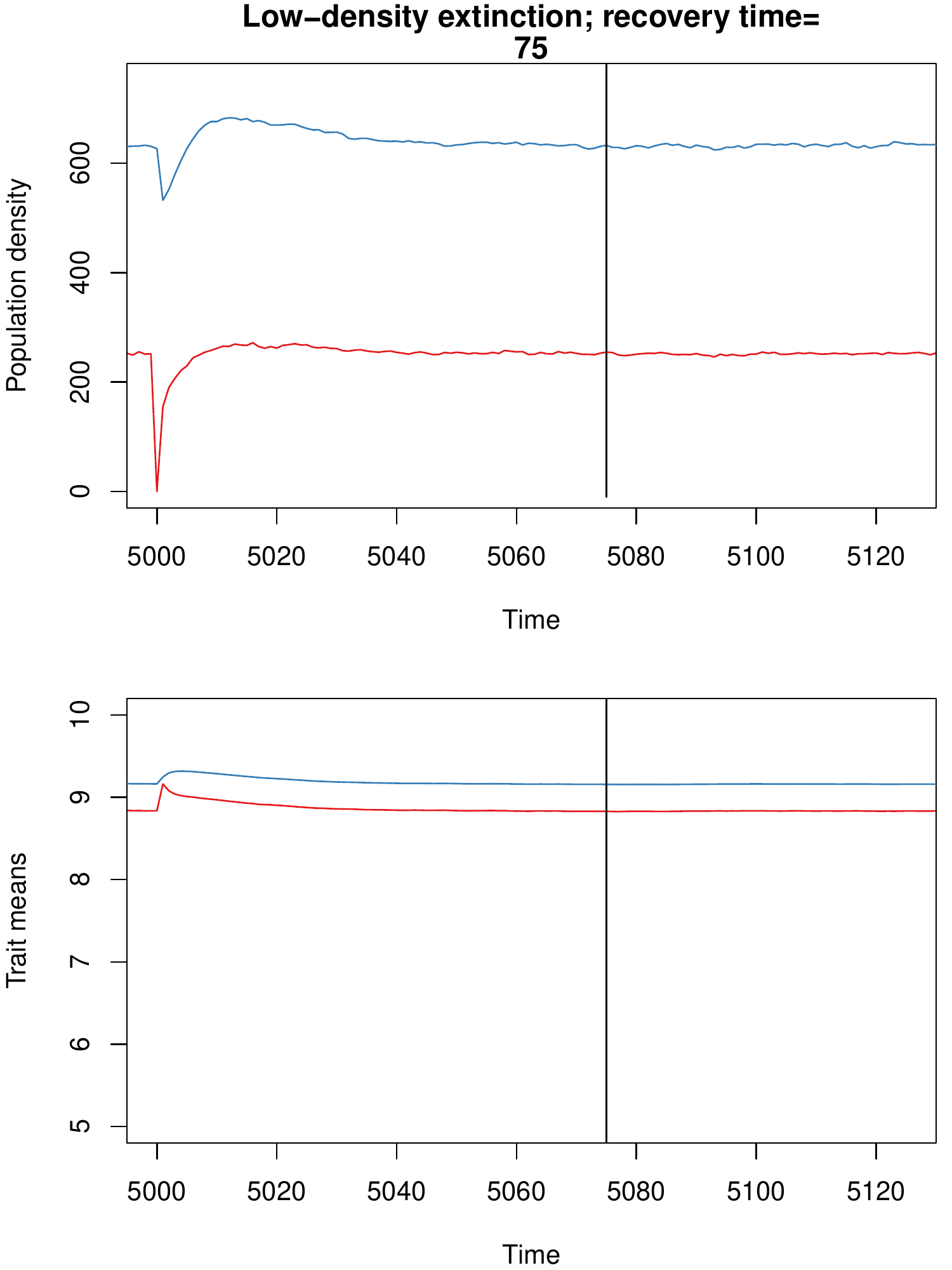}
\caption{
Extinction of low-density population with a high constant straying rate $m=0.4$ and low trait heritability $h^2=0.2$ (see figure \ref{fig:relax}a).
Black line marks the calculated point of recovery post-perturbation.
Trait optima are $\theta_1 = 10$ (blue population trajectory) and $\theta_2 = 5$ (red population).
} \label{fig:relaxtraj_ldlh}
\end{figure*}

Straying had non-linear impacts on the recovery time of populations. 
When the dominant state population (well adapted and high density) goes extinct, high rates of straying allow recolonization of the extirpated habitat and quick recovery (figure \ref{fig:relax}a) because the surviving population has a mean trait value skewed towards the optimum of the affected habitat (figure \ref{fig:relaxtraj_hdlh}).
Yet, as straying decreases, recovery time for the disturbed dominant state population increases, in part because there is time for the trait distribution to move back towards the trait optimum of the subordinate state population.
In contrast, when the subordinate state population (maladapted and low density) is wiped out, recovery rates are fastest at low to intermediate levels of straying.
Because the mean trait values of both populations are skewed towards those of the dominant population, when the subordinate population collapses under high rates of straying, selection against the flood of maladapted individuals that stray into the recovering population extends the length of time required for it to return to its steady-state (figure \ref{fig:relaxtraj_ldlh}).
When both populations are both dramatically reduced, recovery time is generally fastest at lower levels of straying due to reduced mixing of maladaptive phenotypes.
Near the onset of the cusp bifurcation, recovery time increases explosively, however this is -- as the name implies -- characteristic of \emph{slow} dynamics occurring near critical transitions \citep{Scheffer:2009gg,Kuehn:2010p2591}.

Increased rates of straying lowers phenotypic diversity ($\Delta \mu^* = |\mu_i^*-\mu_j^*|$, evaluated at the steady-state) because both local and remote populations are increasingly homogenized.
The loss of phenotypic diversity is greater with increased straying when trait heritability is low because traits take longer to go back to their local optima than they do when heritability is high. 
Hence straying counters the effect of diversifying local adaptation. 
Less intuitively, we observe a discrete jump towards low phenotypic diversity as the cusp bifurcation is crossed (figure \ref{fig:traitdiff}).
Although the development of alternative stable states elevates the portfolio effect due to the variance-dampening effects of the aggregate, entering this dynamic regime also results in a substantial decline in phenotypic diversity, which may have less predictable adverse effects on the population.\\

\noindent{\bf (b) The effects of collective navigation and density-dependent straying} \\
If we assume that the rate of straying is density-dependent, the probability that an individual strays $m_0$ determines the rate of straying within the population, such that $m(t)$ becomes lower as $N(t)$ increases, likely due to the effects of collective decision-making \citep{Berdahl:2016dx} (Eq. \ref{eq:ddm}).
Density dependence alters the straying rate at steady-state population densities because $0 < m^* < m_0$, and this serves to rescale both the strength of the PE as well as the recovery time, but does not change the qualitative nature of our findings.
In the alternative stable state regime, because each population exists at different steady-state densities, there are likewise two alternative straying rates $(m_i^*,m_j^*)$: the higher straying rate is associated with the low-density population, and the lower straying rate is associated with the high-density population.
We assessed metapopulation robustness across a range of $(m_i^*,m_j^*)$ values by varying the probability that an individual strays $m_0$, which is positively and linearly related to $(m_i^*,m_j^*)$.
We find that the portfolio effects generated in systems with density-dependent straying are qualitatively similar to systems with constant straying, however there are some important quantitative differences.
First, the PE associated with the high-density (low $m^*$) population is the same as that for a system with a constant $m$ (figure \ref{fig:thetaPE}a).
As $m^*$ increases, we observe an increase in the PE relative to systems with constant $m$.

\begin{figure*}
  \captionsetup{justification=raggedright,
singlelinecheck=false
}
\centering
\includegraphics[width=0.9\textwidth]{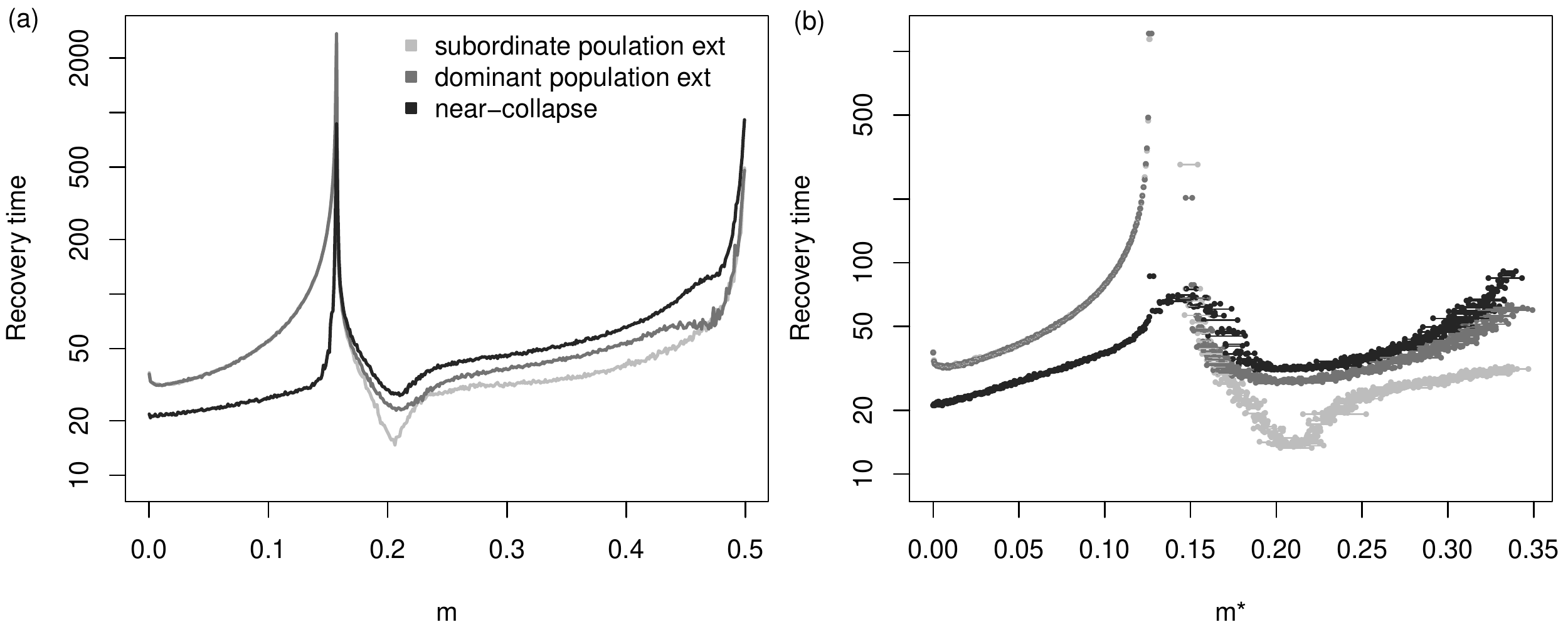}
\caption{
Recovery time of $N_T$ following the extinction of either the low-density (light gray) or high-density (gray) population, or the near-collapse of both (dark gray) assuming (a) constant straying rates $m$ and (b) density-dependent straying rates (evaluated at the steady-state $m^*$) with trait heritability $h^2=0.8$.
If $m$ is density-dependent, in the alternative stable state regime there are two straying rates observed: one each for the low- and high-density populations, respectively, which are linked by a horizontal line.
} \label{fig:relax_highh}
\end{figure*}

Density-dependent straying alters these recovery times (figure \ref{fig:relax}b; note difference in x-axis scale). 
First, in comparison with constant stray rates, density-dependent straying lowered recovery times at elevated straying rates for near-collapse of both populations and extirpation of the subordinate population.
In contrast, at low straying rates, near-collapse of both populations required more time to recover. 
Trait heritability had a large effect on recovery times, with near-collapse requiring a more protracted time to recover in the alternative steady-state regime (figure \ref{fig:relax_highh}).
In general, the lower recovery time for systems with increased $m^*$ mirrors an elevated PE with higher density-dependent straying rates (figure \ref{fig:thetaPE}a).
Together, analysis of both PE and recovery time suggests that although density-dependent straying does not appear to change the `dynamic landscape' in our model, it does appear to promote robustness, particularly in the case of near-collapse of both populations when straying is high.
\\

\begin{figure*}
  \captionsetup{justification=raggedright,
singlelinecheck=false
}
\centering
\includegraphics[width=0.4\textwidth]{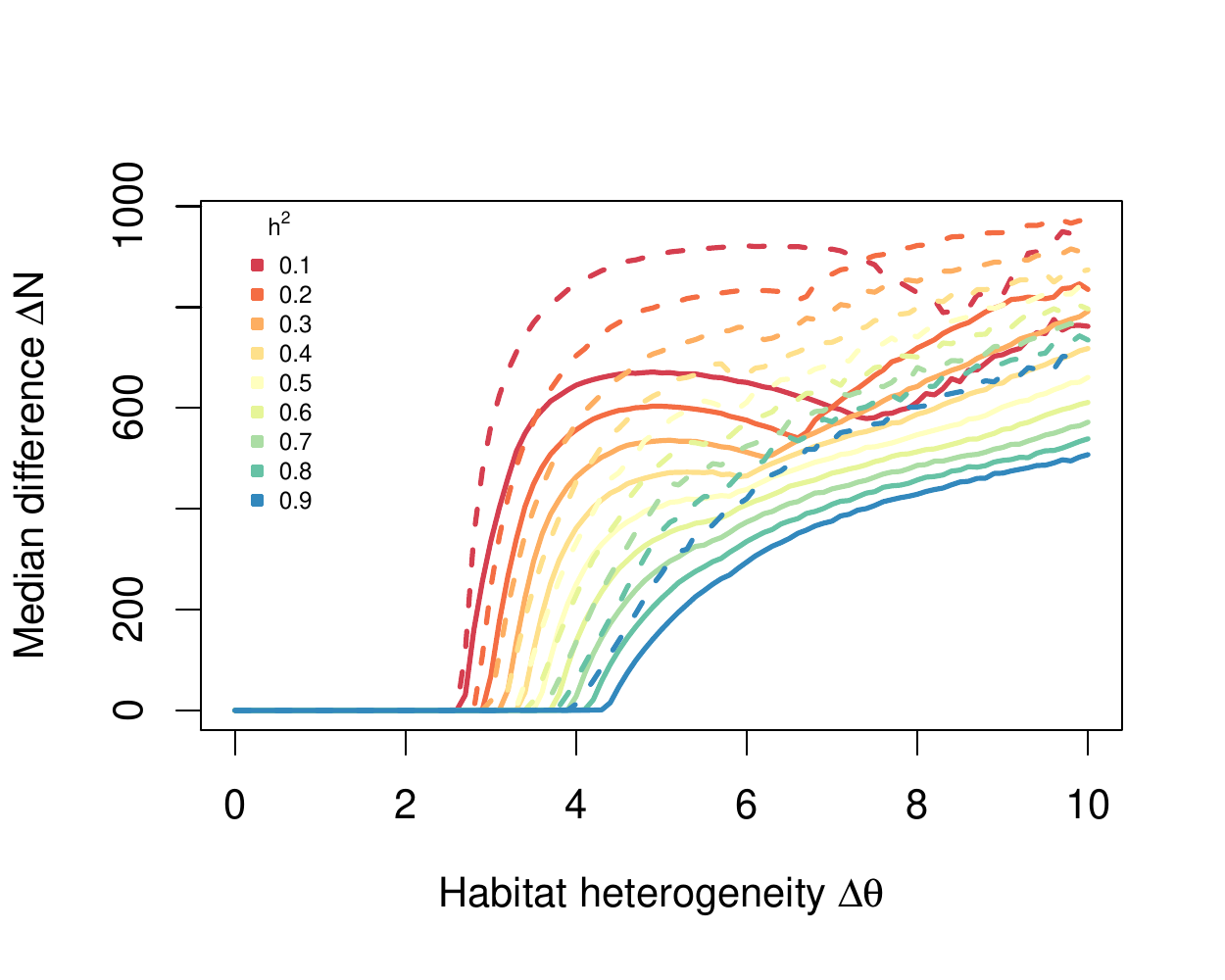}
\caption{
Median difference in population densities taken over the straying rate as a function of habitat heterogeneity $\Delta\theta$.
Solid lines are for constant $m$; dashed lines are for density-dependent $m$.} \label{fig:thetadiffN}
\end{figure*}

\begin{figure*}
  \captionsetup{justification=raggedright,
singlelinecheck=false
}
\centering
\includegraphics[width=0.8\textwidth]{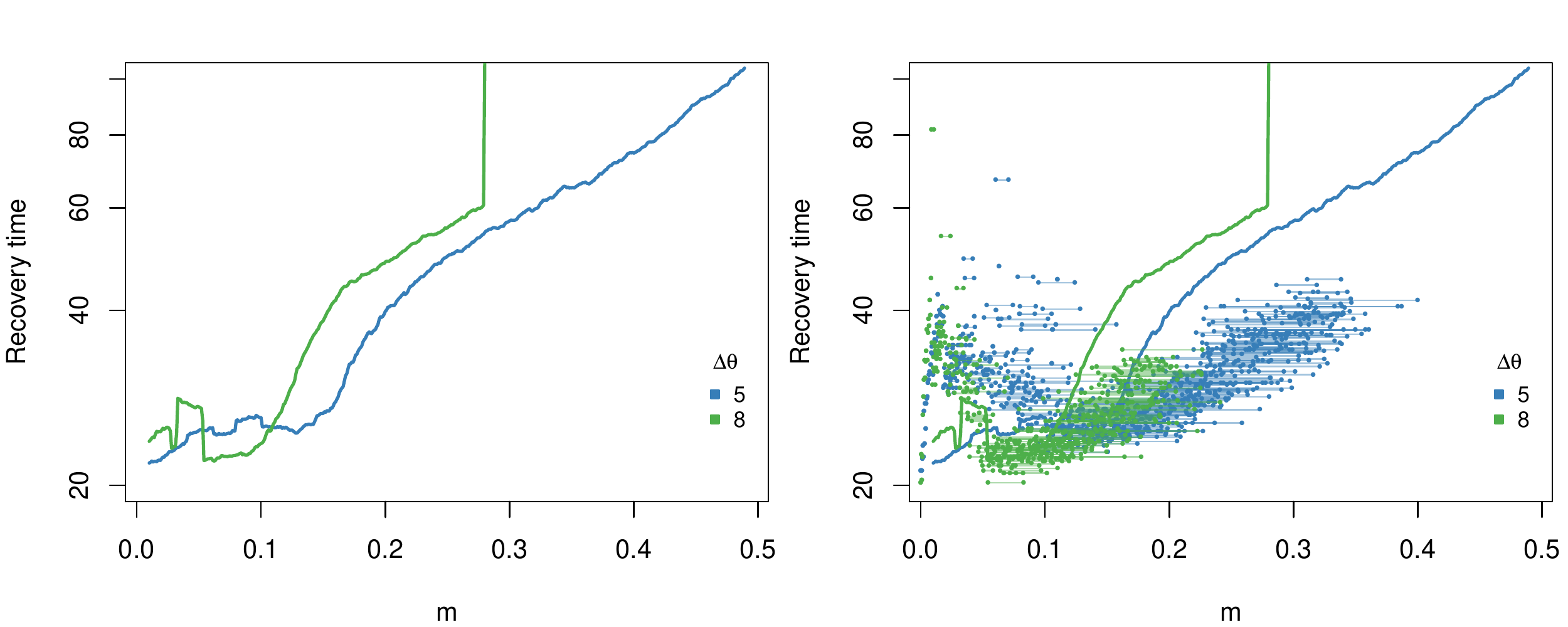}
\caption{
(a) Recovery time after near collapse of both populations as a function of straying rate $m$ and habitat heterogeneity $\Delta\theta$.
(b) The same as (a) but including recovery times when straying is density-dependent evaluated at $m^*$, shown by linked point pairs.
Recovery times for systems with density-dependent straying are longer when straying is low and shorter when straying is high, mirroring the change in portfolio effects with respect to density-dependent straying shown in figure \ref{fig:thetaPE}.
} \label{fig:relaxtheta}
\end{figure*}

\noindent{\bf (c) The role of habitat heterogeneity and changing selective landscapes}\\ 
\noindent With the onset of straying, we find that increasingly divergent trait optima generally lower $N_T^*$ and exaggerate $\Delta N^*$, and this is particularly pronounced for density-dependent straying (figure \ref{fig:thetadiffN}). 
The impact of habitat heterogeneity on the portfolio effect and recovery time is more complex, serving to emphasize the nonlinear relationship between rates of straying and metapopulation robustness. 
As habitat heterogeneity increases, alternative steady-states appear at lower straying rates -- with the crossing of the cusp bifurcation, accompanied by a peak in the PE -- whereas the magnitude of increase in the PE also increases (figure \ref{fig:thetaPE}b), lowering recovery time (figure \ref{fig:relaxtheta}).
For increased rates of straying, greater habitat heterogeneity erodes the PE (figure \ref{fig:thetaPE}b) and increases the recovery times (figure \ref{fig:relaxtheta}).
In tandem, these results suggest that habitat heterogeneity promotes robustness when straying rates are low, and erodes robustness when straying rates are high.

\begin{figure*}
  \captionsetup{justification=raggedright,
singlelinecheck=false
}
  \centering
  \includegraphics[width=0.35\textwidth]{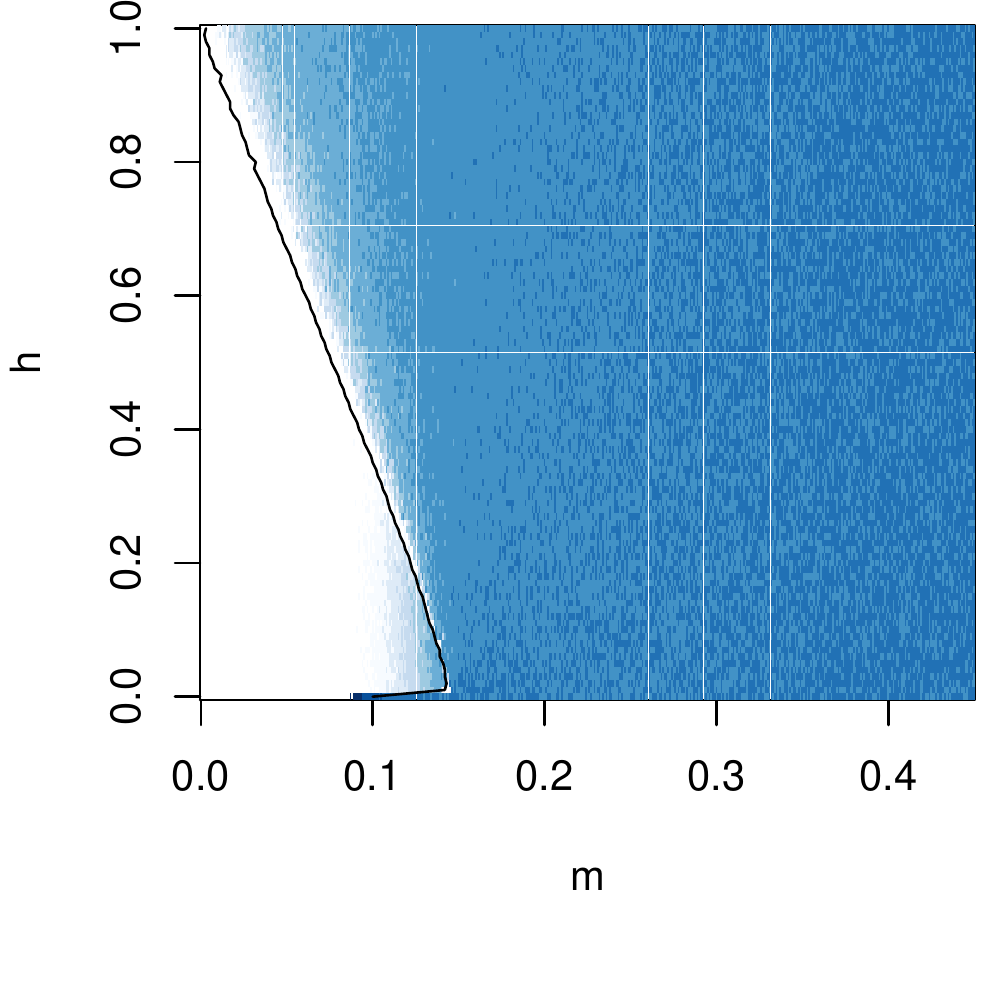}
  \caption{
  Portfolio effects as a function of straying rate $m$ and trait heritability $h^2$ when the rate of straying is $m = 0.5(1 + \Delta\theta)^{-1}$. Alternative steady-states emerge for low values of $m$ (left of the cusp bifurcation, denoted by the black line), whereas a single steady-state exists for high $m$.
  } \label{fig:mthetaPE}
\end{figure*}

Until now, we have treated the rate of straying and habitat heterogeneity as independent parameters, however they may also be assumed to covary.
For instance, if sites are separated by greater distance, they may be assumed to have increased habitat heterogeneity as well as lower rates of straying.
Alternatively, individuals may be genetically predisposed to stray into sites that are more similar, such that greater between-site heterogeneity will correspond to lower straying rates.
We implemented this inverse relationship by setting $m = 0.5(1 + \Delta\theta)^{-1}$ where maximum straying is assumed to occur at $m=0.5$ (perfect mixing; figure \ref{fig:mthetarelation}).
This assumes that $m$ is greater for lower $\Delta\theta$, such that there are low rates of straying between dissimilar (distant) sites and high straying rates between similar (close) sites.
Under these conditions we find that alternative stable states appear for very low rates of straying (figure \ref{fig:mthetaPE}). 
As the straying rate increases and $\Delta\theta$ decreases, a single stable state emerges as the cusp bifurcation is crossed, which is opposite the pattern observed when straying is independent of habitat heterogeneity.

\begin{figure}
  \captionsetup{justification=raggedright,
singlelinecheck=false
}
  \centering
  \includegraphics[width=0.5\textwidth]{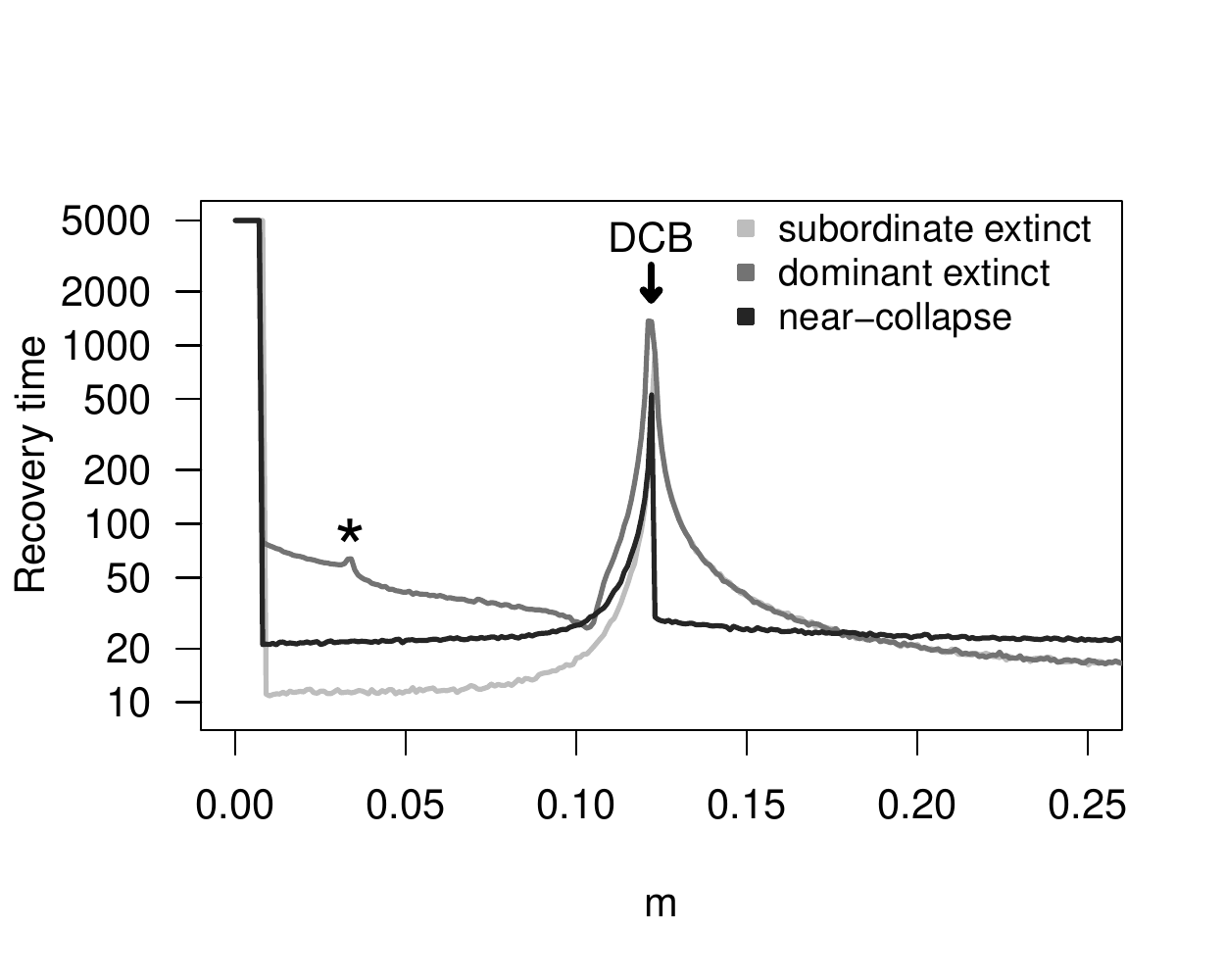}
  \caption{
  Recovery times under three disturbance types for systems where higher habitat heterogeneity $\Delta \theta$ corresponds to lower straying $m$ (see figure \ref{fig:mthetarelation}), and vice versa. 
  The `$*$' marks the value of $m$ below which (low straying between highly heterogeneous environments) there is an inversion between subordinate/dominant states following extinction of the dominant population.
  DCB marks the discrete cusp bifurcation.
  } \label{fig:mtheta}
\end{figure}

\begin{figure*}
  \captionsetup{justification=raggedright,
singlelinecheck=false
}
\centering
\includegraphics[width=0.35\textwidth]{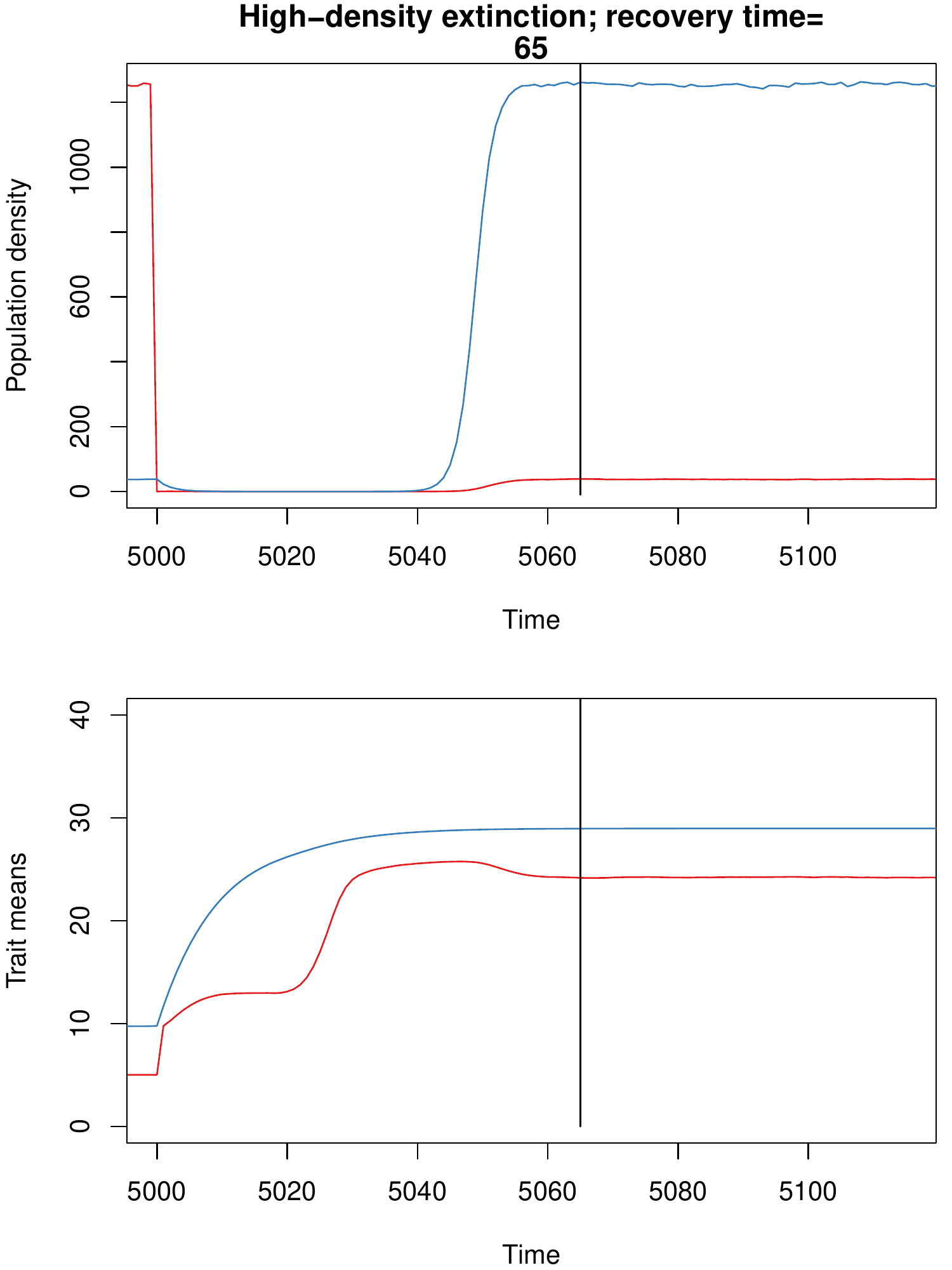}
\caption{
Population inversion where increased differences in trait optima between sites $\Delta\theta$ corresponds to lower rates of straying $m$.
At low rates of straying $m=0.02$ ($\Delta\theta=24$), extinction of the dominant population leads to slower-than-expected recovery times because the subordinate population is isolated enough to evolve towards its own trait optimum. 
In this case, $m$ is less than $m=0.034$ (denoted by the asterisk in figure \ref{fig:mtheta}), such that isolation allows the subordinate population to `run away' from the influence of the dominant population.
This leads to a switch in subordinate/dominant states for the two populations.
If $m$ is low but greater than $0.034$, isolation permits the subordinate population to `run away' from the influence of the dominant population, until it is overwhelmed by the recovering dominant population, and reverts back to its previous trait mean prior to disturbance.
} \label{fig:inertia}
\end{figure*}

\begin{figure*}
  \captionsetup{justification=raggedright,
singlelinecheck=false
}
  \centering
  \includegraphics[width=0.8\textwidth]{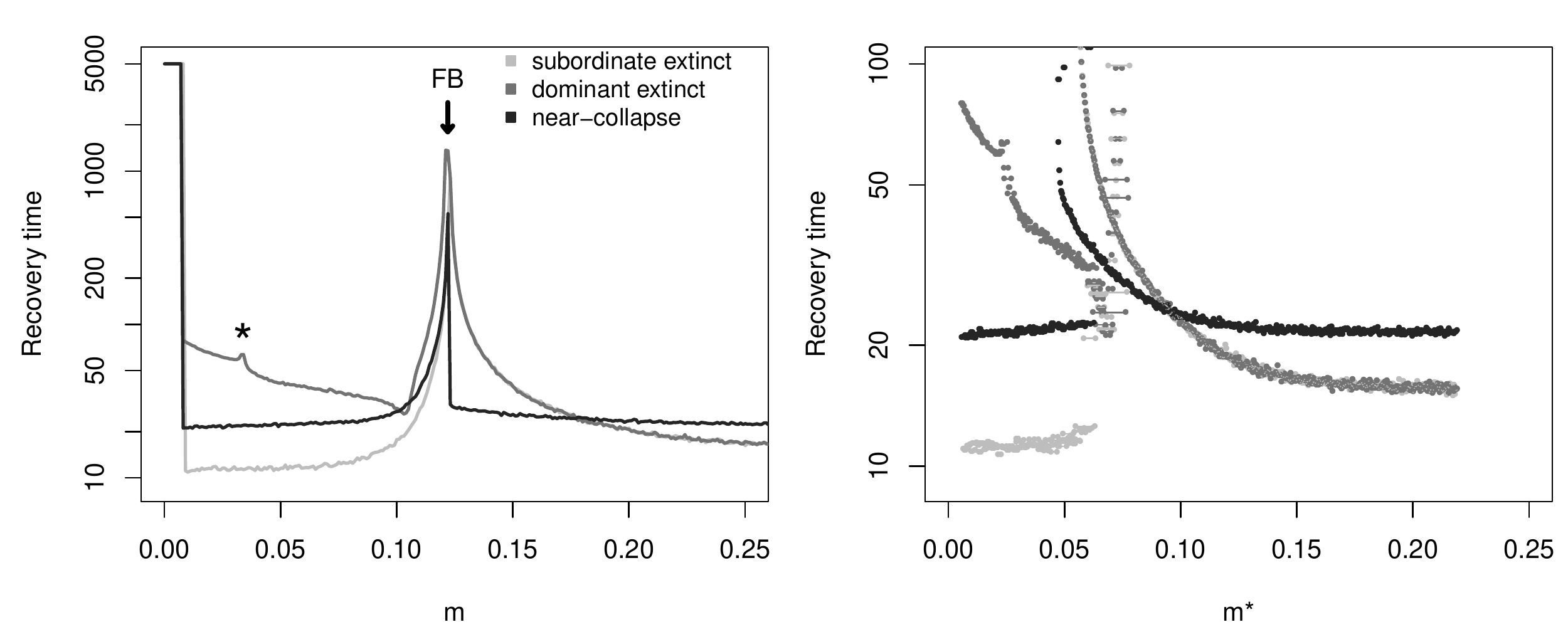}
  \caption{
  Recovery times for three disturbance types when the straying rate covaries with habitat heterogeneity as $m = 0.5(1 + \Delta\theta)^{-1}$ for constant (a) and density-dependent (b) straying rates.
  The cusp bifurcation is not as clear in (b) because $\Delta\theta$ is a function of the individual straying rate $m_0$, whereas the x-axis in (b) is the straying rate at the steady-state $m^*$.
  Despite this difference, the general behavior shown in (a) are the same in (b).
  } \label{fig:mthetamvm}
\end{figure*}

There are two notable dynamics that emerge following extinction of the dominant population at low rates of straying between dissimilar (high $\Delta\theta$) sites (figure \ref{fig:mtheta}).
(\emph{i}) Above a threshold $m$ value, the dominant population recovers quickly enough that the evolving subordinate phenotype is overwhelmed by incoming strays, and it shifts back to its pre-disturbance subordinate state;
(\emph{ii}) below a threshold $m$ value, there is an \emph{inversion} between subordinate and dominant states: because there is enough time and isolation for the subordinate trait mean to shift towards its local optimum, and away from that of the recovering dominant population, the dominant population becomes subordinate, and the subordinate population becomes dominant (figure \ref{fig:inertia}).
This threshold value of $m$, below which the inversion dynamic behavior occurs, is marked by the asterisk in figure \ref{fig:mtheta}, and holds for both constant and density-dependent straying (figure \ref{fig:mthetamvm}).

\section{Discussion}

We have shown that density-dependent straying between populations consistent with collective navigation, coupled with localized selection against donor phenotypes, has a large and nonlinear impact on dynamic properties that affect metapopulation robustness.
We measured robustness as:
1) the average-CV portfolio effect \citep{Anderson:2013gb,Anonymous:2015gf}, a statistical metric commonly used to assess the buffering capacity of metapopulations, and
2) the recovery time, defined here as the time required for the aggregate metapopulation biomass $N_T^*$ to return to its steady-state following an induced disturbance, which is mechanistically linked to persistence \citep{Ovaskainen:2002il}.
In our eco-evolutionary model of dispersal and natural selection between two populations, we show that these statistical and mechanistic descriptors of metapopulation dynamics and robustness are tightly coupled (figure \ref{fig:PE}d), which is not uncommon for diverse metrics of stability \citep{Donohue:2013iu}.
Taken as a whole, our results point to an important role of density-dependent straying in the colonization and recovery dynamics within metapopulations, while also underscoring the risk of straying by individuals with maladaptive traits to reduce the productivity of locally adapted stock complexes.

\begin{figure*}
  \captionsetup{justification=raggedright,
singlelinecheck=false
}
\centering
\includegraphics[width=0.35\textwidth]{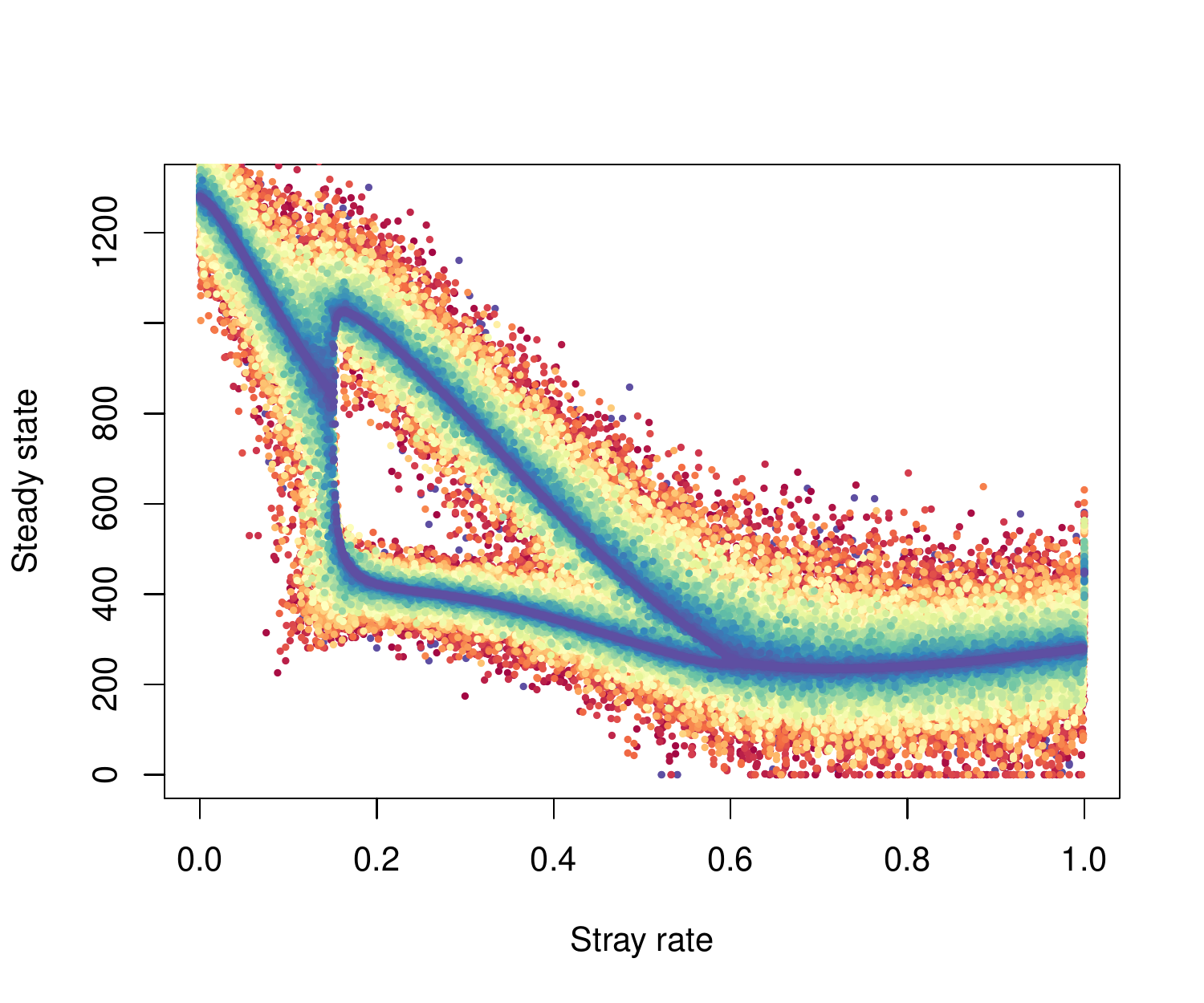}
\caption{
Steady-state densities of both populations as a function of $m$, where a cusp bifurcation indicates the emergence of alternative steady-states: one in a dominant state and one in a subordinate state.
Steady-states for populations with symmetrical values ($\alpha=0$) in the vital rates $r_{\rm max}$ and $\beta$ are shown with cool tones.
As the asymmetry among populations between sites increases ($\alpha>0$), their vital rates diverge, such that the maximal growth at sites 1 and 2 is now $r_{\rm max}(1)=r_{\rm max}(1+\tilde{rv}_1)$ and $r_{\rm max}(2)=r_{\rm max}(1+\tilde{rv}_2)$ where $rv_{1,2}$ are independently drawn from $\rm{Normal}(0,\alpha)$ and $r_{\rm max}=2$. 
Similarly the strength of density dependence is calculated at sites 1 and 2 as $\beta(1)=\beta(1+\tilde{rv}_1)$ and $\beta(2)=\beta(1+\tilde{rv}_2)$ where $\tilde{rv}_{1,2}$ are independently drawn from $\rm{Normal}(0,\alpha)$ and $\beta=0.001$.
Steady-states for populations with increasingly asymmetric values ($\alpha\rightarrow 0.1$) for vital rates $r_{\rm max}$ and $\beta$ are shown in warmer tones.
} \label{fig:symmetry}
\end{figure*}

A salient finding from our results is that straying can lead to the emergence of alternative stable states, pushing one of the populations to high density (the \emph{dominant state}), and one to low density (the \emph{subordinate state}).
This pattern has been observed in other models of eco-evolutionary dynamics with explicit space \citep{Ronce:2001dp}, suggesting that it may be a general feature of spatially-linked populations that evolve toward local optima while being hindered by dispersal.
An important aspect of our framework is that there are similar forces that dictate interactions within and between sites, and this naturally results in a symmetry that could be perceived as limiting the relevance of our findings for natural (and inherently less symmetric) systems.
Although the emergence of alternative stable states via a cusp bifurcation is characteristic of symmetrical systems, we find that increasing the asymmetry in the vital rates of populations across sites does not significantly alter the presence or position of the DCB (figure \ref{fig:symmetry}).
That these patterns arise in alternative formulations and are relatively insensitive to parameter asymmetry suggests that the dynamical features observed have potentially widespread ramifications for the evolutionary dynamics of spatially connected populations in general.
<

%

An intermediate straying rate increases metapopulation robustness. 
Results from our model reveal that the presence of just enough straying to cause formation of alternative stable states increases the portfolio effect (figure \ref{fig:PE}a). 
We note that we do not consider the extremely high PE at the DCB, matched by an extremely long recovery time, to be an indicator of robustness, as the dynamics exactly at or very close to a bifurcation are unlikely to be realized in nature.
Previous theoretical work has shown that increased connectivity may erode portfolio effects in herring metapopulations, where straying is also thought to be density-dependent \citep{Secor:2009ena}.
Although high levels of dispersal in our system generally supports this finding, the interplay between dispersal and PE is more subtle when selection for local adaptations is considered.
Low to intermediate levels of straying result in an elevated PE, increasing the buffering capacity of the metapopulation.

Although PE is measured at the steady-state, low to intermediate rates of straying also appear to have a beneficial effect on transient dynamics.
When there is just enough straying to cause alternative stable states, the time to recovery following an induced disturbance declines, though -- as with the PE -- it then grows if straying becomes large (figure \ref{fig:relax}a).
In the alternative steady-state regime, a lower rate of straying inhibits admixture of maladapted individuals.
Following a large disturbance, such as the near-collapse of both dominant and subordinate populations, this limited mixing increases the growth rates of both populations, permitting faster recovery times. 
If the rate of straying becomes too high, an influx of maladapted individuals into both populations inhibits local growth rates, and recovery slows.

\begin{figure*}
  \captionsetup{justification=raggedright,
singlelinecheck=false
}
\centering
\includegraphics[width=0.35\textwidth]{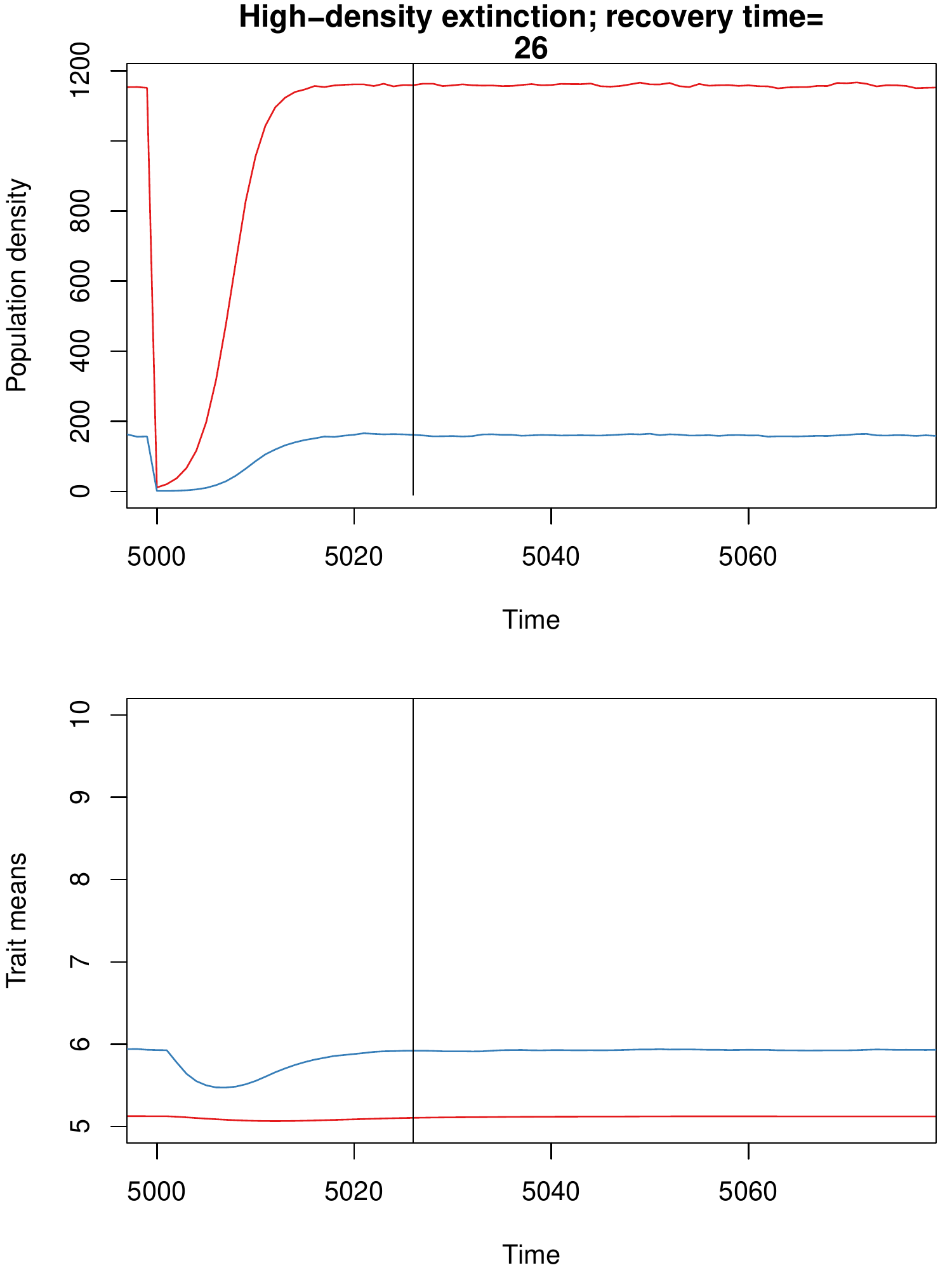}
\caption{
Near collapse of both populations with a low straying rate $m=0.1$ and low trait heritability $h^2=0.2$ (see figure \ref{fig:relax}a).
Black line marks the calculated point of recovery post-perturbation.
Trait optima are $\theta_1 = 10$ (blue population trajectory) and $\theta_2 = 5$ (red population).
} \label{fig:relaxtraj_bothlh}
\end{figure*}

This themed issue formalizes the role of collective movement in the ecology of natural systems and illuminates a signature of collective navigation in animal populations on the move.
We highlight three important findings that contribute to our understanding of collective movement suggesting that density-dependent straying may play an important role in the persistence of metapopulations over evolutionary time.
First, the inclusion of density-dependent straying does not qualitatively alter either (\emph{i}) steady-state or (\emph{ii}) transient dynamics of our eco-evolutionary model, but effectively rescales measures of robustness to the lower straying rates that emerge as a consequence of the coupled dynamics.
Second, compared to systems with constant dispersal, density-dependent straying appears to increase the portfolio effect across a range of straying rates (figure \ref{fig:thetaPE}a). 
Third, density-dependent straying reduces the time to recovery following disturbance, and this is particularly true in the case of near-collapse of the metapopulation (figure \ref{fig:relax}b).
In the case of near-collapse, although both populations inherit low population densities, the mean trait values of both are skewed towards the optimum of the dominant population.
Due to density-dependent straying, low population densities lead to greater dispersal, and while this increased connectivity primarily facilities the growth of the dominant population (because the trait means are closer to the dominant optimum), because the dominant population contains the bulk of the aggregate biomass, the overall recovery time is lessened significantly (figures \ref{fig:relax}b, \ref{fig:relaxtraj_bothlh}).


Salmon are distributed and stray across a diverse range of habitats, and the rates of straying between geographically diverse sites can be plastic and idiosyncratic \citep{Westley:2015to}.
Our surrogate measure for habitat heterogeneity is the difference in trait optima between sites $\Delta\theta$.
In general, our findings indicate that increased habitat heterogeneity promotes robustness (higher PE, shorter time to recovery) when straying rates are low, but may erode robustness when straying rates are high (figure \ref{fig:thetaPE}b, solid lines).
This may be particularly consequential for populations that are spatially adjacent but separated by sharp environmental boundaries, such that trait optima are divergent yet dispersal is relatively high.
Such a scenario plays out repeatedly in the context of wild and hatchery-produced salmon. 
Although wild and hatchery populations may occur close on the landscape, and indeed often are sympatric within the same river network, the selective environments to which they are locally adapted differ dramatically \citep{Christie:2012bj}. 
Straying of domesticated hatchery-produced fish from release sites and spawning in the wild drastically reduces the productivity of wild populations through competition and outbreeding depression \citep{Chilcote:2003bb,Araki:2007cm}.

\begin{figure*}
  \captionsetup{justification=raggedright,
singlelinecheck=false
}
\centering
\includegraphics[width=0.35\textwidth]{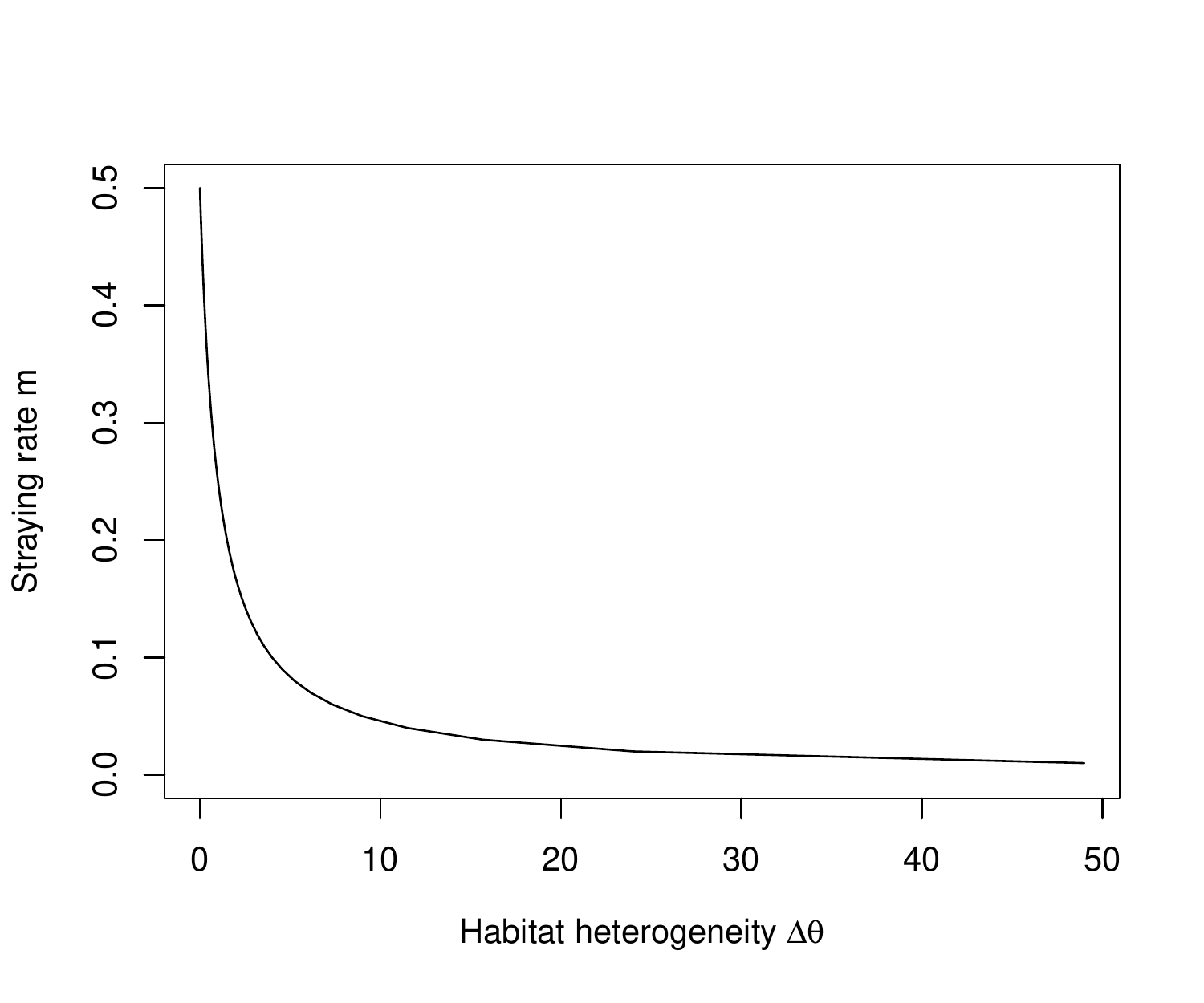}
\caption{
In some cases, habitat heterogeneity may be assumed to determine the rate of straying, if for example:
1) sites are distributed over greater spatial distances, where habitat differences are assumed to be greater between more distant sites, or 2) individuals have behaviors promoting dispersal between habitats that are more similar. To examine such cases, we use the relationship $m= 0.5(1 + \Delta\theta)^{-1}$ where maximum straying is assumed to occur at $m=0.5$ (perfect mixing).
} \label{fig:mthetarelation}
\end{figure*}

In other cases, habitats that are closer in space can be assumed to have greater similarity in environmental conditions than those that are geographically distant, and phenotypes of more proximately located populations should be more similar \citep{Reisenbichler:1988ex,Fraser:2011co,Westley:2012ui}.
It is thus reasonable to expect a larger number of straying individuals between sites that are geographically proximate and indeed evidence corroborates this prediction \citep{Candy:2000hu,JPE:JPE1383}.
Alternatively, salmon that cue to specific environmental conditions may be more likely to stray into sites that are structurally and physiognamically more similar \citep{Peterson:2014gy}.
These considerations justify imposing a direct relationship between the rate of straying $m$ and habitat heterogeneity: as site dissimilarity increases, so too should the straying rate decrease (figure \ref{fig:mthetarelation}).
When habitat heterogeneity and the rate of straying are linked in this way, we show that very small amounts of either constant or density-dependent straying result in long recovery times for the dominant population because there is time for selection to push the subordinate trait mean away from the optimum of the dominant population (figure \ref{fig:mtheta}). 
Such a dynamic is accompanied by an inversion in the alternative stable states following the disturbance, resulting in a state shift in dominance.
Thus, management activities that alter dispersal rates by outplanting individuals or reconnecting disconnected habitats could have unintended eco-evolutionary consequences  \citep{Anderson:2013bf,Pess:2014isa}.

Although our study was inspired by salmon metapopulations, the results have general implications for the conservation and management of other migratory metapopulations as well. 
Because changes in straying rates can have large and nonlinear impacts on robustness, human activities that alter straying rates could have unintended consequences. 
For example, salmon produced by hatcheries often stray into proximate wild populations \citep{Brenner:2012gl}, and these recipient populations can have lower fitness due in part to the introduction of maladapted genes \citep{Ford:2002ip}. 
We show that there is an intermediate straying rate where disturbed populations that are recovering by the introduction of maladapted strays recover fastest: if the straying rate is too low or too high, recovery times increase (figure \ref{fig:relax}).
This finding suggests that salmon stocking efforts that aim to lower recovery times following dam removal could actually prolong recovery if the rate at which individuals are introduced and the suitability of those fish in that habitat (i.e. their measure of pre-adaptation) is not taken into account.
Ongoing examinations of experimental restocking in the recently re-opened Elwha River (Washington State) will provide empirical insight into the potential short- and long-term consequences of facilitated recovery \citep{Liermann:2017gj}. 


The portfolio effect and the time to recovery following a disturbance are independent and correlated measures of metapopulation robustness that take into account both steady-state and transient dynamics.
We show that these measures of robustness are strongly influenced by the rate at which individuals from donor populations stray into habitats occupied by recipient populations. 
Importantly, density-dependent straying, which may occur when individuals collectively navigate, can both increase the portfolio effect and lower the time to recovery following a disturbance, which is anticipated to promote persistence. 
Therefore, preserving the biological processes that facilitate this collective behavior may be an important conservation target in its own right, echoing the sentiments of  Hardesty-Moore et al. \citep{HardestyMoore:wg}. 
We suggest that understanding the spatial complexity of metapopulations dispersing across heterogeneous environments, in tandem with the mosaic of selective forces acting on those environments, may be key to uncovering those factors that promote persistence in the wild.
\\ \\

%
%

\noindent {\bf Competing interests:} The authors declare no competing interests
\\
\noindent {\bf Author contributions:} JDY and JWM conceived of the initial project design. JDY and JPG designed the modeling framework and conducted the analyses. JDY, JPG, PAHW, and JWM interpreted the results, and drafted and wrote the manuscript.
\\
\noindent {\bf Data Accessibility:} Code is made available at https://github.com/jdyeakel/SalmonStrays
\\
\noindent {\bf Acknowledgements:} We thank Sean Anderson for helpful discussions and comments on the manuscript. We also thank the guest editors Andrew Berdahl, Dora Biro, and Colin Torney, for inviting us to contribute to this themed issue, and two anonymous reviewers for their insightful comments and critiques. J.D.Y. was supported by startup funds at the University of California, Merced and an Omidyar Postdoctoral Fellowship at the Santa Fe Institute. J.P.G. was supported by a James S. McDonnell Foundation Postdoctoral Fellowship in Complex Systems at the University of California, Merced. P.A.H.W. was supported by the UA Foundation at the University of Alaska Fairbanks. J.W.M. was supported by the Liber Ero Research Chair in Coastal Science and Management at Simon Fraser University.


\clearpage



\clearpage

\beginsupplement

\end{document}